\documentclass[prd,showpcs,amsmath,amssymb,nofootinbib,longbibliography,twocolumn,showpacs,notitlepage]{revtex4-1}
\usepackage{multirow}
\usepackage{epsfig}
\usepackage{amsmath}
\usepackage{bm}
\usepackage{times}
\usepackage{graphicx}
\usepackage{color}
\usepackage{slashed}
\usepackage{graphicx}
\usepackage{amsmath}
\usepackage[latin1]{inputenc}
\usepackage{hyperref}
\usepackage{soul}

\def\bea{\begin{eqnarray}} 
\def\eea{\end{eqnarray}}
\def\bean{\begin{equation*}}
\def\eean{\end{equation*}} 
\def\nn{\nonumber}
\def\beaal{\begin{align}}
\def\eeaal{\end{align}}

\begin{document}
 
\title{Left-Right SU(4) Vector Leptoquark Model for Flavor Anomalies}

\author{Bartosz~Fornal}
\affiliation{Department of Physics, University of California, San Diego, \\ 9500 Gilman Drive, La Jolla, CA 92093, USA \vspace{1mm}}
\author{Sri~Aditya~Gadam}
\affiliation{Department of Physics, University of California, San Diego, \\ 9500 Gilman Drive, La Jolla, CA 92093, USA \vspace{1mm}}
\author{Benjam\'{i}n~Grinstein \vspace{1mm}}
\affiliation{Department of Physics, University of California, San Diego, \\ 9500 Gilman Drive, La Jolla, CA 92093, USA \vspace{1mm}}
\date{\today}

\begin{abstract}
Building on our recent proposal to explain the experimental hints of
new physics in $B$ meson decays within the framework of Pati-Salam
quark-lepton unification, through the interactions of the $(3,1)_{2/3}$
vector leptoquark, we construct a realistic model of this type based
on the gauge group ${\rm SU}(4)_L \times {\rm SU}(4)_R \times {\rm
  SU}(2)_L \times {\rm U}(1)'$ and consistent with all experimental
constraints. The key feature of the model is that ${\rm SU}(4)_R$ is
broken at a high scale, which suppresses right-handed lepton flavor
changing currents at the low scale and evades the stringent bounds
from searches for lepton flavor violation. The mass of the leptoquark
can be as low as $10 \ {\rm TeV}$ without the need to introduce
mixing of quarks or leptons with new vector-like
fermions. We provide a comprehensive list of model-independent
  bounds from low energy processes on the couplings in
  the effective Hamiltonian that arises from generic leptoquark
  interactions, and then apply these to the model presented here. We discuss various meson decay channels that can be used to probe the model and we investigate the prospects for discovering the new gauge boson at future colliders.\vspace{6mm}
\end{abstract}

\maketitle

\section{Introduction}\label{1} 
The Standard Model (SM) provides a remarkably successful description
of nature at the elementary particle level and, so far, there
are only a handful of experimental indications of deviations
from its predictions. Perhaps the most significant direct hint of
physics beyond the SM are the recently observed anomalies in
$B$ meson decays \cite{Aaij:2014ora,Aaij:2017vbb}, which
suggest that lepton universality might be violated. Assuming that
those anomalies are not a result of experimental systematics, they are
best accounted for by the vector leptoquark $(3,1)_{2/3}$ or
$(3,3)_{2/3}$ \cite{Alonso:2015sja,Alonso:2014csa,Kosnik:2012dj}. However, building viable UV
complete models involving those particles is challenging,
especially in  light of very stringent constraints on lepton flavor
violation (LFV) from various experimental searches.
\vspace{1mm}

The first attempt to construct a vector leptoquark model for the $R_{K^{(*)}}$ anomalies was made in  \cite{Assad:2017iib}, where we proposed that the vector leptoquark $(3,1)_{2/3}$ explaining the anomalies might be the gauge boson of a theory with Pati-Salam unification. The conclusion was that
the minimal model based on ${\rm SU}(4) \times {\rm SU}(2)_L \times
{\rm SU}(2)_R$ is not capable of this because  of strict bounds on
kaon and $B$ meson rare decays
\cite{Valencia:1994cj,Smirnov:2007hv,Smirnov:2008zzb,Carpentier:2010ue,Kuznetsov:2012ai,Smirnov:2018ske}. The
  underlying problem  in that model arises from the interference between
  left-handed (LH) and right-handed (RH) lepton flavor changing currents. We
outlined a possible solution to this: extending the gauge group to ${\rm
  SU}(4)_L \times {\rm SU}(4)_R \times {\rm SU}(2)_L \times {\rm
  U}(1)'$ and breaking ${\rm SU}(4)_R$ at a high scale, such that the
RH lepton flavor changing currents are suppressed. 
\vspace{1mm}

A viable realization of this idea is the subject of this paper.\\ We demonstrate  that a Pati-Salam gauge leptoquark as light as $10 \ {\rm TeV}$ can explain the $R_{K^{(*)}}$ anomalies and remain consistent with all experimental bounds without introducing any mixing of quarks and leptons  with new fermions. We discuss in detail the constraints arising from LFV searches and show that the absence of RH lepton flavor changing currents  relaxes the bounds considerably. The model is expected to have clean signatures at future colliders, which we  investigate in the case of the prospective $100 \ {\rm TeV}$ machine.
\vspace{1mm}

 Several
other models for the flavor anomalies based on Pati-Salam unification
have been proposed, some appearing almost immediately after our
  initial work
\cite{DiLuzio:2017vat,Calibbi:2017qbu,Bordone:2017bld,Barbieri:2017tuq,Blanke:2018sro,Greljo:2018tuh,new}. Those
models overcome the experimental constraints by mixing all or a subset
of SM quarks and leptons with new vector-like fermions. 
Other approaches to account for the $B$ meson decay
  anomalies   involving scalar leptoquarks or $Z'$
rather than vector leptoquarks  have been also proposed  (see, e.g. \cite{Marzocca:2018wcf,Aydemir:2018cbb,Becirevic:2018afm,Guadagnoli:2018ojc,Li:2018rax,Faber:2018qon,Heeck:2018ntp,Allanach:2018lvl}).
\vspace{1mm}

In App.\,\ref{aapp1}
we provide a model-independent analysis of  the low energy consequences of a
$(3,1)_{2/3}$ vector
leptoquark \linebreak
that interacts with both LH and  RH fields. We
present  an extensive list of bounds from flavor physics on
generic coupling constants in this model-independent  approach; 
App.\,\ref{aapp1}  is thus a resource in its own right, of  use to
researchers interested in any specific model of this type. Appendix\,\ref{aapp2} is one such example, where 
 we apply the results of App.\,\ref{aapp1} to the
specific Pati-Salam model  constructed  in this work.  The
calculations in App.\,\ref{aapp1} update and extend previous results
\cite{Valencia:1994cj,Smirnov:2007hv,Smirnov:2008zzb,Carpentier:2010ue,Kuznetsov:2012ai,Smirnov:2018ske}.
For instance, for $B$ decays we use the most recent lattice results for the form factors \cite{Bouchard:2013pna}, which weaken the bounds considerably compared to assuming  the nonphysical values $f_+=f_0=1$  adopted previously in the literature.\vspace{-1mm}

\section{The model}\label{2}
The theory we propose is based on the gauge group
\bea\label{gaugegroup}
{\rm SU}(4)_L \times {\rm SU}(4)_R \times {\rm SU}(2)_L \times {\rm U}(1)' \ .
\eea
The crucial feature of the model is that the subgroup ${\rm SU}(4)_R$ is broken at a much higher scale than ${\rm SU}(4)_L$, leading to a suppression of RH lepton flavor changing currents.

\vspace{3mm}
\noindent
\centerline
{\bf {\emph{Fermion particle  content}}}\vspace{1mm}\\
The matter fields in the model, along with their decomposition into ${\rm SU}(3)_c \times {\rm SU}(2)_L \times {\rm U}(1)_Y$ multiplets, are
\bea \label{fermions}
\hat\Psi_{L} &=& (4,1, 2, 0) \,=\, (3,2)_{\frac16} \oplus (1,2)_{-\frac12}\,,\nn\\[2pt]
\hat\Psi_{R}^u&=&(1,4,1, \tfrac12) \,=\,   (3,1)_{\frac23} \oplus  (1,1)_0\,,\nn\\[2pt]
\hat\Psi_{R}^d&=&(1,4,1, -\tfrac12) \,=\, (3,1)_{-\frac13} \oplus   (1,1)_{-1} \ ,
\eea

\
\vspace{-6mm}
\bea
\hat\chi_{L}&=&(\bar4,1,2, 0) \,=\,  (\bar3,2)_{-\frac16} \oplus (1,2)_{\frac12}\,,\nn\\[2pt]
\hat\chi_{R}&=&(1,\bar4,2, 0) \,=\,   (\bar3,2)_{-\frac16} \oplus (1,2)_{\frac12}\,\nn 
\eea
for each generation, where $\hat\Psi_{L} $, $\hat\Psi_{R}^u$ , $\hat\Psi_{R}^d$ contain the SM fields $Q_L$, $L_L$, $u_R$, $d_R$, $e_R$ and a RH neutrino $\nu_R$, whereas $\hat\chi_{L}$, $\hat\chi_{R}$ assure gauge anomaly cancellation and result in two vector-like pairs of fields ${Q'}_{\!\!L}$, ${Q'}_{\!\!R}$ and ${L'}_{\!\!L}$, ${L'}_{\!\!R}$ that are heavy and do not mix with SM fermions. This is the minimal fermion content for a consistent theory based on the gauge group (\ref{gaugegroup}).

\vspace{4mm}
\noindent
\centerline
{\bf {\emph{Scalar sector and symmetry breaking}}}\vspace{1mm}\\
The Higgs sector contains the scalar representations
\bea
&&\hat\Sigma_L = \,(4, 1, 1, \tfrac12)\,, \ \ \ \ \hat\Sigma_R = \,(1, 4, 1, \tfrac12)\,, \ \ \ \ \hat\Sigma = (\bar4, 4, 1, 0)\, ,\nn\\
&&\hspace{8mm}\hat{H}_{d} \,= \,(4, \bar4, 2, \tfrac{1}{2}) \, , \ \ \ \ \hat{H}_{u} = (4, \bar4, 2, -\tfrac{1}{2})\, .
\eea
The scalar potential is given in App.\,{\ref{aap3}}.
The  parameters can be chosen such that the fields  $\hat\Sigma_L$, $\hat\Sigma_R$ and $\hat{\Sigma}$ develop the vacuum expectation values (vevs),
\bea \label{vevs}
\langle \hat\Sigma_L\rangle = \frac{v_L}{\sqrt2}\!
\begin{pmatrix}
\,0\,\\
0\\
0\\
1
\end{pmatrix} , \ \ \ \ \langle \hat\Sigma_R\rangle =\frac{v_R}{\sqrt2}\!
\begin{pmatrix}
\,0\,\\
0\\
0\\
1
\end{pmatrix} , \nn \\
\langle \hat\Sigma\rangle = \frac{v_\Sigma}{\sqrt2}\!
\begin{pmatrix}
\,1 & 0& 0& 0\\
0 & 1& 0& 0\\
0 & 0& 1& 0\\
0 & 0& 0& z\,
\end{pmatrix} , \ \ \ \ \ \ \ \ \ \ \ \ \ \ \ \ \ 
\eea
where $z>0$.
This results in the symmetry breaking pattern
\bea\label{SSB}
\begin{aligned}
&{\rm SU}(4)_L\times {\rm SU}(4)_R \times {\rm SU}(2)_L \times {\rm U}(1)'\\[2pt]
& \ \ \ \rightarrow \ {\rm SU}(3)_c \times {\rm SU}(2)_L \times {\rm U}(1)_Y \ .
\end{aligned}
\eea
The relation between the SM hypercharge $Y$ and the ${\rm U}(1)'$ charge $Y'$ is given by
\bea
Y = Y' + \sqrt{\frac23} \left({T}_L^{15} + {T}_R^{15}\right)\,,
\eea
where

\vspace{-7mm} 
\bea
T^{15}_L = {T}^{15}_R= \frac{1}{2\sqrt6}\,{\rm diag}(1,1,1,-3) \ . 
\eea
The scalar representations decompose into SM fields as
\bea\label{scalars}
\begin{aligned}
\hat\Sigma_{L} &=(3,1)_{\frac23} \oplus (1,1)_{0}\ , \ \ 
\hat\Sigma_{R} =(3,1)_{\frac23} \oplus (1,1)_{0}\ ,\\
\hat\Sigma \ \, &= (8,1)_0 \,\oplus (3,1)_{\frac23}\oplus (\bar3,1)_{-\frac23} \oplus 2\,(1,1)_{0} \ ,\\
\hat{H}_d &=  (8, 2)_{\frac12}  \oplus (3, 2)_{\frac76} \oplus (\bar3,2)_{-\frac16} \oplus 2\,(1,2)_{\frac12} \\
&\equiv  O_1 \oplus T_{1} \oplus T_2^\dagger \oplus S_1 \oplus S_2 \ ,\\
\hat{H}_u &=  (8, 2)_{-\frac12}  \oplus (3, 2)_{\frac16} \oplus (\bar3,2)_{-\frac76} \oplus 2\,(1,2)_{-\frac12} \\
&\equiv   O_2 \oplus T_{3} \oplus T_4^\dagger \oplus S_3^* \oplus S_4^* \ .
\end{aligned}
\eea
Under the symmetry breaking pattern (\ref{SSB}) the $\hat{H}_{d}$, $\hat{H}_{u}$ fields have $(\bar4,4) \rightarrow (\bar3 \oplus 1) \otimes(3 \oplus 1)$; 
$S_1$, $S_3$ stand for the singlets in $1\otimes 1$, while $S_2$, $S_4$ are the singlets in $\bar3 \otimes 3$.
The components of $\hat\Sigma_{R}$, $\hat\Sigma_{L}$, $\hat\Sigma$ have masses on the order of the ${\rm SU}(4)_R$ and ${\rm SU}(4)_L$ breaking scales. 
This is also the natural mass scale for the components of $\hat{H}_d$ and $\hat{H}_u$. However, as shown in App.\,{\ref{aap4}}, it is possible to fine-tune the parameters of the potential such that only one linear combination of the fields $S_{1,2,3,4}$  is light. In particular, there exists a choice of parameters for which the light state is given by
\bea\label{expa}
H = - \,c_e\, S_1 - c_d\,S_2 + c_\nu\, S_3 + c_u\, S_4 \ ,
\eea
where $c_u \approx 1 \gg c_d \gg c_\nu$ and $ 1 \gg c_e \gg c_\nu$, with the ratio $c_d:c_e \approx m_b:m_\tau $.
This reduces the scalar sector of the model to that of the SM  at low energies.

\vspace{4mm}
\noindent
\centerline
{\bf {\emph{Gauge sector}}}\vspace{1mm}\\
The gauge and kinetic terms are
\begin{align}\label{lagrr}
\mathcal{L}_{\rm g+k}=&  -\tfrac{1}{4} G^A_{L\mu\nu} G^{A  \mu\nu}_L \!-\tfrac{1}{4} G^A_{R\mu\nu} G^{A  \mu\nu}_R \! -\tfrac{1}{4} W^a_{\mu\nu} W^{a  \mu\nu} \nonumber\\
& -  \tfrac{1}{4} Y_{\mu\nu}' {Y'}^{\mu\nu}\!+  |D_\mu \hat\Sigma_L|^2+ |D_\mu \hat\Sigma_R|^2 \ \  \nonumber\\
& + |D_\mu \hat\Sigma|^2 +|D_\mu \hat{H}_d|^2 + |D_\mu \hat{H}_u|^2\ \  \nonumber\\
&+   \overline{\hat\Psi}_{L} i\slashed{D}\, {\hat\Psi_{L}} + \overline{\hat\Psi} {}^u_{R}\, i\slashed{D} \,{\hat\Psi_{R}^u} + \overline{\hat\Psi} {}_{R}^d \,i \slashed{D} \,{\hat\Psi_{R}^d} \ ,
\end{align}
with  $A = 1, ..., 15$ and $a=1,2,3$. 
The gauge covariant derivative takes the form
\bea
\begin{aligned}
D_\mu =  & \ \, \partial_\mu + i g_L \hspace{0.4mm}G_{L\mu}^A T_L^A +  i g_R \hspace{0.4mm}G_{R\mu}^A T_R^A \\
&+ i g_2 \,W_\mu^a \,t^a + i g_{1}' \,Y'_{\mu}\, Y' \ ,
\end{aligned}
\eea
where $T_L^A$, $T_R^A$, $t^a$, $Y'$ are  the ${\rm SU}(4)_L$, ${\rm SU}(4)_R$, ${\rm SU}(2)_L$, ${\rm U}(1)'$ generators. The gauge couplings at the low scale are related to the SM strong and hypercharge couplings via
\bea\label{gauge}
g_s =\! \frac{g_L\, g_R}{\sqrt{g_L^2 + g_R^2}}\ , \ \ \ g_1 \!=\! \frac{g'_1g_L g_R}{\sqrt{\tfrac23{g'_1}^{\!2}(g_R^2+g_L^2)+g_L^2 \,g_R^2}} \ . \ \ \ \ \ \ \ 
\eea
The new gauge bosons are
\bea
&&X_L = (3,1)_{\frac23} \, , \  \ \ X_R = (3,1)_{\frac23}\, , \ \ \ \ 
G' = (8,1)_0 \, , \ \ \ \nn\\
&&\hspace{12mm}\,Z'_L = (1,1)_0\, , \ \ \ \  Z'_R = (1,1)_0 \ .
\eea
The mass of $G'$ is $M_{G'} = \tfrac{1}{\sqrt2}\sqrt{g_L^2+g_R^2}\, v_{\Sigma}$. The squared mass matrix for the gauge leptoquarks $X_L$, $X_R$ is 
\bea\label{pm}
\mathcal{M}^2_{X} =\tfrac14\!
\begin{pmatrix}
\,g_L^2\big[v_L^2+v_\Sigma^2(1+z^2)\big]  & -\,2\,g_L g_R v_\Sigma^2 z   \\[9pt]
-\,2\,g_L g_R v_\Sigma^2 z & g_R^2\big[v_R^2 + v_{\Sigma}^2(1+z^2)\big]\,
\end{pmatrix} \!. \nn\\
\eea
The leptoquark mass eigenstates can be written as
\bea\label{xmatr}
\begin{pmatrix}
X_1     \\
X_2
\end{pmatrix}
= 
\begin{pmatrix}
\cos{\theta_{4}} &    \sin{\theta_{4}}  \\
-\sin{\theta_{4}} &   \  \cos{\theta_{4}}  
\end{pmatrix}
\begin{pmatrix}
X_L      \\
X_R
\end{pmatrix},
\eea
where the mixing angle $\theta_4$ depends on the parameters in Eq.\,(\ref{pm}).
In the limit $v_R \gg v_L$ and  $v_R \gg v_\Sigma$ the mixing vanishes, $\sin\theta_4 = 0$, and the leptoquark masses become
\bea
M_{X_1} &=& \tfrac12\, g_L \sqrt{v_L^2+v_\Sigma^2(1+z^2)} \ , \nn\\
M_{X_2} &=& \tfrac12 \,g_R \, v_R \ .
\eea
The $Z'_L$ and $Z'_R$  squared masses are given by  the two nonzero eigenvalues of the matrix
\bea\label{MZ}
&&{\mathcal{M}}_{Z'}^2 = \tfrac{3}{8} \times \\
&&\!\begin{pmatrix}
 g_L^2\!\left[v_L^2\!+\!v_\Sigma^2(\frac13\!+\!z^2)\right]  & -g_L g_R v_\Sigma^2(\frac13\!+\!z^2)   &-{\frac{\sqrt2}{\sqrt3}}{g'_1g_L }v_L^2  \\[9pt]
\!-g_L g_R v_\Sigma^2(\frac13\!+\!z^2)  &\!\!g_R^2\!\left[v_R^2\!+\!v_\Sigma^2(\frac13\!+\!z^2)\right]\! &-{\frac{\sqrt2}{\sqrt3}}{g'_1g_R }v_R^2\\[9pt]
-{\frac{\sqrt2}{\sqrt3}}{g'_1g_L}v_L^2&-{\frac{\sqrt2}{\sqrt3}}{g'_1g_R}v_R^2 &\frac23{g'_1}^{\!2} (v_L^2+v_R^2)
\end{pmatrix}\!. \nonumber
\eea

\vspace{1mm}
\noindent
Taking the limit $v_R \gg v_L$ and $v_R\gg v_\Sigma$ yields
\bea
&&M_{Z'_L} =\sqrt{\frac{ {g'_1}^{\!2}(g_L^2+g_R^2)+\frac32\,g_L^2g_R^2}{8({g'_1}^{\!2}\!+\frac32g_R^2)}} \sqrt{3v_L^2 + v_\Sigma^2(1+3z^2)} \ ,\nn\\
&&M_{Z'_R} = \frac12\sqrt{ {g'_1}^{\!2}\!+\tfrac32 g_R^2 } \ v_R \ .
\eea

\vspace{2mm} 
\noindent
\centerline
{\bf {\emph{Fermion masses}}}\vspace{1mm}\\
The Yukawa interactions are
\bea\label{lll}
\mathcal{L}_{Y}&= &y_{ij}^{d}\,\overline{\hat\Psi^{i}_L} \hat{H}_d \hat\Psi_{R}^{d\hspace{0.2mm}j}   +y_{ij}^{u}\,\overline{\hat\Psi^i_L} \hat{H}_u \hat\Psi_{R}^{u j}
+ Y_{ij}\,\overline{\hat\chi^{i}_L} \,\hat{\Sigma} \,\hat\chi_{R}^{j}  + {\rm h.c.}\ \nn\\[6pt]
&\supset& \,   y_{ij}^{d}\,\overline{L^i_L} {S}_1  e_{R}^{j} + y_{ij}^{d}\,\overline{Q^i_L} {S}_2 d_{R}^{j}  + y_{ij}^{u}\,\overline{L^i_L} {S}_3^*  \nu_{R}^{j}  \nn \\
&+&\, y_{ij}^{u}\,\overline{Q^i_L} {S}_4^*  u_{R}^{j}   + \tfrac{1}{\sqrt2}Y_{ij}v_\Sigma\big(\overline{{Q'}^{i}_{\!\!L}} \, {Q'}_{\!\!R}^{j} + z\,\overline{{L'}^{i}_{\!\!L}} \, {L'}_{\!\!R}^{j}\big) + {\rm h.c.} \nn\\[4pt]
&\supset&    - \,c_e \,y_{ij}^{d}\,\overline{L^i_L} H  e_{R}^{j} -c_d \,y_{ij}^{d}\,\overline{Q^i_L} H d_{R}^{j}  + c_\nu\,y_{ij}^{u}\,\overline{L^i_L} \tilde{H}  \nu_{R}^{j} \nn  \\
&&+\, c_u\,y_{ij}^{u}\,\overline{Q^i_L} \tilde{H}   u_{R}^{j}   + \tfrac{1}{\sqrt2}Y_{ij}v_\Sigma\big(\overline{{Q'}^{i}_{\!\!L}} \, {Q'}_{\!\!R}^{j} + z\,\overline{{L'}^{i}_{\!\!L}} \, {L'}_{\!\!R}^{j}\big)  \nn\\
&& +\,  {\rm h.c.} \ ,
\eea
where $i,j\!=\!1,2,3$ are family indices and the coefficients ``$c$'' are those in Eq.\,(\ref{expa}).
Typically, in theories with quark-lepton unification,  the  up-type quark and neutrino masses of a given generation are the same at the unification scale, and similarly the down-type quark and charged lepton masses. In our model this is not the case, but since there  are only two Yukawa matrices $y^u$ and $y^d$,  without additional mass contributions the hierarchy of the up-type quark masses is, a priori,  the same as for the neutrinos, and the down-type quark mass hierarchy  the same as for the charged leptons at the unification scale.
\vspace{1mm}

Regarding the up-type quarks and neutrinos, for which the experimentally determined  mass hierarchies differ considerably, this is solved by introducing a new scalar representation
$
\hat{\Phi}_{10} = (1, \overline{10}, 1, -1)$. 
If the 
SM singlet component of $\hat{\Phi}_{10}$ develops a vev $v_{10}$ at a high scale, this provides a seesaw mechanism for the neutrinos via the interaction
\bea
y_{ij}^{u\hspace{0.2mm}\prime}\, \overline{(\hat\Psi_{R}^{u i})^{c}}\,\hat{\Phi}_{10}\hat\Psi_{R}^{uj}  .
\eea
The contribution to the up-type quarks  vanishes. Therefore, the up-type quark masses are  $m_u \sim y^u v$, whereas the neutrino  masses are $m_{\nu} \sim (c_\nu \,y^u v)^2 / (y^{u\hspace{0.2mm}\prime}v_{10})$.
\vspace{1mm}

The relative mass hierarchies of the down-type quarks versus charged leptons are not in vast disagreement with experiment.
The running of the masses will largely account for $m_b/m_\tau$. One can also introduce the scalar representation $\hat{\Phi}_{15} = (15, 1, 1, 0)$ into the model, with the SM singlet component developing the vev $v_{15} \,{\rm diag}(1,1,1,-3)$. New mass contributions to the down-type quarks and charged leptons would then result from  loop processes, parameterized via the effective dimension five interaction $\,{ {y}_{ij}^{d\hspace{0.2mm}\prime}}\,\overline{\hat\Psi^{i}_L} \hat{H}_d \hat\Psi_{R}^{d\hspace{0.2mm}j} \hat{\Phi}_{15}/\Lambda$, and mediated, e.g., by heavy vector-like fermions, leading to additional mass splitting.

\vspace{3mm}
\noindent
\centerline
{\bf {\emph{Flavor structure}}}\vspace{1mm}\\
In terms of SM fermion mass eigenstates, the 
interactions of the vector leptoquarks with quarks and leptons are given by
\bea\label{19}
\mathcal{L} \,& \supset & \frac{g_L}{\sqrt2}\, X_{L \mu} \Big[L^u_{ij}\,(\bar u^i\gamma^\mu P_L\,\nu^j)+L^d_{ij}\,(\bar d^i \gamma^\mu P_L \,e^j) \Big]\\
&+& \frac{g_R}{\sqrt2}\, X_{R\mu} \Big[R^u_{ij}\,(\bar u^i\gamma^\mu P_R\,\nu^j)+R^d_{ij}\,(\bar d^i \gamma^\mu P_R\, e^j) \Big]  \!+  {\rm h.c.}\, ,\nn
\eea
where $L^{u/d}$, $R^{\,u/d}$ are unitary mixing matrices. They are related via $L^u = V L^d U$ and $R^u = V R^d U$, where $V$ is the Cabibbo-Kobayashi-Maskawa  matrix and $U$ is the Pontecorvo-Maki-Nakagawa-Sakata matrix.

\vspace{3mm}
\noindent
\centerline
{\bf {\emph{Proton stability}}}\vspace{1mm}\\
The vector boson $(3,1)_{2/3}$ does not mediate proton decay \cite{Assad:2017iib} and neither do any of the scalars in our model. In particular, for the scalar $(3,2)_{1/6}$, which by itself would be problematic \cite{Arnold:2012sd}, gauge invariance forbids  tree-level proton decay.  
In broader terms, the Lagrangian in Eq.~(\ref{lll}) is invariant under the global symmetries ${\rm U}(1)'_{B}$ and ${\rm U}(1)'_{L}$, with the matter fields $\Psi_L$, $\Psi^d_R$ and $\Psi^u_R$ carrying charges $B'=L'=1/4$ and all scalar fields being neutral.  After symmetry breaking the charges under the remaining global ${\rm U}(1)_{B}$ and ${\rm U}(1)_{L}$ are 
\bea
\begin{aligned}
B &= B' +\tfrac{1}{\sqrt6} \left({T}_L^{15} + {T}_R^{15}\right) ,\\
L &= L' -{\tfrac{\sqrt6}{2}} \left({T}_L^{15} + {T}_R^{15}\right) ,
\end{aligned}
\eea
which are  simply the SM baryon and lepton number. Proton decay is thus forbidden at all orders in perturbation theory.

\section{Flavor anomalies}
In this section we discuss how the vector leptoquark of ${\rm SU}(4)_L$ can explain the recent hints of physics beyond the SM in $B$ meson decays, i.e., 
the deficit in the ratios
\bea\label{RRstar}
\begin{aligned}
R_K &= \frac{{\rm Br}\big(B^+ \to K^+ \mu^+\mu^-\big)}{{\rm Br}\big(B^+ \to K^+ e^+e^-\big)}\ ,\\
R_{K^*} \!&= \frac{{\rm Br}\big({B^0} \to {K^{*0}}\mu^+\mu^-\big)}{{\rm Br}\big({B^0} \to {K^{*0}} \,e^+e^-\big)}   \ 
\end{aligned}
\eea
with respect to SM predictions  \cite{Aaij:2014ora,Aaij:2017vbb}. For an analysis of the anomalies at the effective operator level see  \cite{Geng:2017svp,Ciuchini:2017mik,Hiller:2017bzc,DAmico:2017mtc,Altmannshofer:2017yso,Capdevila:2017bsm,Celis:2017doq,Alok:2017sui}. \vspace{1mm}

To describe  the decays in Eq.~(\ref{RRstar}) quantitatively,  it is convenient to start out from the effective Lagrangian for flavor changing neutral current processes with a $b\to s$ transition. Up to four-quark operators, it can be written as
\bea\label{sop}
\mathcal{L} = \ && \frac{4G_F}{\sqrt2} V_{tb}V_{ts}^* \sum_{i,j}\Big[\,\sum_{k=7}^{10}C_k^{ij} {\mathcal{O}}_k^{ij(\prime)} +C_\nu^{ij}{\mathcal{O}}_\nu^{ij} \ \ \ \ \ \nn \\
&&+ \ C_S^{ij(\prime)} {\mathcal{O}}_S^{ij(\prime)} + C_P^{ij(\prime)} {\mathcal{O}}_P^{ij(\prime)} \Big]  \ .
\eea
The operators $\mathcal{O}_7^{ij}$ and $\mathcal{O}_8^{ij}$ correspond to electromagnetic and chromomagnetic moment transitions; the $\mathcal{O}_9^{ij(\prime)}$, $\mathcal{O}_{10}^{ij(\prime)}$, $\mathcal{O}_\nu^{kl}$ are the  semileptonic operators
\bea
\begin{aligned}
\mathcal{O}_{9(10)}^{ij}&= \frac{e^2}{16\pi^2} \left(\bar{s}\,\gamma_\mu P_L \,b\right)\big[\,\overline{l^i}\,\gamma^\mu (\gamma_5)\,l^j\,\big]\ ,\\
\mathcal{O}_{9(10)}^{ij\,\prime} &= \frac{e^2}{16\pi^2} \left(\bar{s}\,\gamma_\mu P_R\, b\right)\big[\,\overline{l^i}\,\gamma^\mu (\gamma_5)\,l^j\,\big]\ ,\\
\mathcal{O}_{\nu}^{ij} &= \frac{e^2}{8\pi^2} \left(\bar{s}\,\gamma_\mu P_L \,b\right)\left(\bar{\nu}^i\gamma^\mu P_L \nu^j\right) ;
\end{aligned}
\eea
and \,${\mathcal{O}}_S^{ij(\prime)}$,  ${\mathcal{O}}_P^{ij(\prime)}$ are the scalar operators
\bea
\begin{aligned}
{\mathcal{O}}_S^{ij(\prime)} &= \frac{e^2}{16\pi^2} \big[\bar{s} \,P_{R(L)} b\big]\big(\,\overline{l^i}\,l^j\big) ,\\
{\mathcal{O}}_P^{ij(\prime)} &= \frac{e^2}{16\pi^2} \big[\bar{s} \,P_{R(L)} b\big]\big(\,\overline{l^i}\gamma_5 \,l^j\big) .
\end{aligned}
\eea
Tensor operators were neglected since they cannot arise from short-distance new physics with SM linearly realized~\cite{Alonso:2014csa}. 

Global fits to the  $R_{K^{(*)}}$ anomalies and other $b \to s \,\ell \,\ell$ data have been performed \cite{Geng:2017svp,Ciuchini:2017mik,Hiller:2017bzc,DAmico:2017mtc,Altmannshofer:2017yso,Capdevila:2017bsm,Celis:2017doq,Alok:2017sui}. These analyses yield similar best fit values for the Wilson coefficients. In what follows we adopt the results of \cite{Geng:2017svp}, i.e.,
\bea\label{c9}
{\rm Re}\left(\Delta C_9^{\mu\mu} -\Delta C_9^{ee}\right) = -{\rm Re}\left(\Delta C_{10}^{\mu\mu}- \Delta C_{10}^{ee}\right)\,\approx - \,0.6 \ , \ \ \ \nonumber\\
\eea
with the contributions to the remaining Wilson coefficients being small.
In our model, the vector leptoquarks $X_1$, $X_2$ modify the coefficients by
\bea
&&\Delta C_9^{ij} = -\Delta C_{10}^{ij} = -\frac{\sqrt2\,\pi^2g_L^2 \,L^d_{2i}L^{d*}_{3j}}{G_F \,e^2\,V_{tb}V_{ts}^*}\!\left[\frac{\cos^2\!\theta_4}{M_{X_1}^2}+ \frac{\sin^2\!\theta_4}{M_{X_2}^2}\right] \!,\nn\\
&&\Delta C_9^{ij\prime}  = \Delta C_{10}^{ij\prime}  = -\frac{\sqrt2\,\pi^2g_R^2 R^d_{2i}R^{d*}_{3j}}{G_F \,e^2\,V_{tb}V_{ts}^*}\!\left[\frac{\sin^2\!\theta_4}{M_{X_1}^2}+ \frac{\cos^2\!\theta_4}{M_{X_2}^2}\right] \!,\nn\\ 
&&\Delta C_S^{ij} = -\Delta C_P^{ij} \nn\\
&&\hspace{9mm}= -\frac{\sqrt2\,\pi^2g_Lg_R R^d_{2i}L^{d*}_{3j}\sin2\theta_4}{G_F \,e^2\,V_{tb}V_{ts}^*} \!\left[\frac{1}{M_{X_1}^2}- \frac{1}{M_{X_2}^2}\right] \!,\nn\\
&&\Delta C_S^{ij\prime} = \Delta C_P^{ij\prime} \nn\\
&&\hspace{10mm}=  -\frac{\sqrt2\,\pi^2g_Lg_R L^d_{2i}R^{d*}_{3j}\sin2\theta_4}{G_F \,e^2\,V_{tb}V_{ts}^*}  \!\left[\frac{1}{M_{X_1}^2}- \frac{1}{M_{X_2}^2}\right] \!,\nn\\[3pt]
&&\Delta C_\nu^{ij} =0 \ .
\eea
Guided by the tightness of the bounds from LFV searches (discussed in Sec.~\ref{3} and App.~\ref{aapp1}), we assume that ${\rm SU}(4)_R$ is broken at a much higher scale than ${\rm SU}(4)_L$, i.e.
\bea
v_R \gg v_L \ \ {\rm and} \ \ v_R \gg v_\Sigma \ .
\eea
This suppresses RH lepton flavor changing currents and results in the contributions to the Wilson coefficients other than $\Delta C_{9,10}^{ij}$ being small. The condition in Eq.~(\ref{c9}) becomes
\bea\label{lmass1}
\frac{M_{X_L}}{g_L \sqrt{{\rm Re}\left(L_{22}^{d}L_{32}^{d*} - L_{21}^{d}L_{31}^{d*}\right)}}\approx 23 \ \rm{TeV} \ .
\eea

\section{Experimental constraints}\label{3}

The leptoquark masses and the mixing matrices are subject to experimental constraints  from a number of null searches for LFV, with the most stringent bounds coming from rare decays of pions \cite{Britton:1992pg,Britton:1993cj,Czapek:1993kc}, kaons \cite{Ambrose:1998cc,Ambrose:1998us,Ambrose:2000gj,Appel:2000tc,Sher:2005sp,Ambrosino:2009aa}, $B$ mesons \cite{Aubert:2006vb,Aubert:2007mm,Aubert:2007rn,Aubert:2008cu,Aaltonen:2009vr,Aaij:2017cza,Aaij:2017vad,Aaij:2017xqt},
 $\tau$ leptons  \cite{Aubert:2006cz,Miyazaki:2007jp,Miyazaki:2010qb,Miyazaki:2011xe} and  $\mu - e$ conversion \cite{Bertl:2006up}.
 \vspace{1mm}
 
Implications of those constraints for Pati-Salam unification have been considered in the literature \cite{Valencia:1994cj,Smirnov:2007hv,Smirnov:2008zzb,Carpentier:2010ue,Kuznetsov:2012ai,Smirnov:2018ske}, but focused on models in which the vector leptoquark $(3,1)_{2/3}$ couples to both LH and RH fermion fields with similar strength. The conclusion of those analyses, updated with the most recent experimental bounds \cite{Aaij:2017cza,Aaij:2017vad,Aaij:2017xqt}, is that the leptoquark mass has to be $\gtrsim 90 \ {\rm TeV}$ \cite{Smirnov:2018ske}. In addition, constraints from searches for $\mu\to e\, \gamma$ when both LH and RH leptoquark interactions are present can push this limit much higher  due to the bottom quark mass enhancement of the one-loop diagram (see App.~\ref{aapp1} and also \cite{Arnold:2013cva} for a discussion of a similar effect in scalar leptoquark models). Such a heavy leptoquark would not explain the $R_{K^{(*)}}$ anomalies, since the required relation, analogous to the one  in Eq.~(\ref{lmass1}), could not be satisfied for a perturbative gauge coupling and unitary mixing matrices.
 \vspace{1mm}

In our model, for a sufficiently high scale of ${\rm SU}(4)_R$ breaking, the constraints arising from the presence of leptoquark RH couplings to fermions are eliminated and the remaining bounds on LH interactions can be satisfied for a significantly lower leptoquark mass. The tightest limits are listed in the Appendix, for arbitrary LH and RH leptoquark interactions in App.~\ref{aapp1} and for the case of just LH interactions in App.~\ref{aapp2}. 
 \vspace{1mm}

If the mixing matrix entries  $L^d_{11}$, $L^d_{12}$ are $\mathcal{O}(1)$, the  limits from searches for $K^0_L \to e^\pm\mu^\mp$ and $\mu\!-\!e$ conversion  a priori push the leptoquark mass up to hundreds of TeV in our model (thousands of TeV for models in which both LH and RH leptoquark interactions are present, due to the enhancement of the scalar current contribution; see App.~\ref{aapp1}).  The bounds, however, are satisfied for a much lighter leptoquark
provided $L^d_{11}, L^d_{12} \ll 1$. Unitarity then implies that $L^d_{13}\approx 1$ and $L^d_{23}, L^d_{33} \ll 1$; therefore  $L^d$  takes the form
\bea\label{matricesL}
L^d \,  \approx {e^{i\phi}}\left( \  
\begin{matrix} \vspace{0.5mm}
\!\delta_1 & \delta_2 & \ 1\\ \vspace{1mm}
\!e^{i\phi_1}\cos\theta&e^{i \phi_2}\sin\theta& \ \delta_3 \\ 
\!-e^{-i\phi_2} \sin\theta&e^{-i\phi_1} \cos\theta& \ \delta_4
 \end{matrix} \ \ \right) ,  
 \eea
 where $|\delta_i| \ll 1$. Note that  the suppression of RH flavor changing currents in our model implies that there are no significant bounds from 
 \,$\pi^0 \to \nu\bar\nu$ \,or\, $K^0_L \to \nu\bar\nu$. 
  \vspace{1mm}

The remaining entries of $L^d$ are subject to further constraints, mainly  from $B$ meson and $\tau$
decays. If both LH and RH leptoquark interactions were present, the $B^0\to \mu^+\mu^-$ decay would
provide the most stringent bound.  However, with only  LH interactions the tightest limits arise
from searches for $B^+ \to K^+ e^\pm\mu^\mp$. We calculated the corresponding branching fractions
(see App.~\ref{aapp1}) using  the most recent lattice results  for the form factors
\cite{Bouchard:2013pna} based on the Bourrely-Caprini-Lellouch parameterization
  \cite{Bourrely:2008za}, which relaxes the bounds considerably compared to taking the
nonphysical values $f_+=f_0=1$ \cite{Kuznetsov:2012ai}. 

The resulting bound on $M_{X_L}$  is minimized for $\theta \approx \pi/4$ and requires merely $M_{X_L}/g_L \gtrsim 9.2 \ {\rm TeV}$. Given the relation between the gauge couplings in Eq.~({\ref{gauge}})  and assuming $g_R\approx \sqrt{3\pi}$ (close to the  perturbative limit) implies $g_L \approx 1.06\,g_s$, where $g_s \approx 0.96$ is the strong coupling constant at $10 \ {\rm TeV}$. This leads to the constraint 
\bea\label{12}
M_{X_L} \gtrsim 10 \ {\rm TeV} \ .
\eea
(If one chose instead $g_L = g_R = \sqrt2 \,g_s$, this would result in the constraint $M_{X_L} \gtrsim 14 \ {\rm TeV} $.)
Saturating the bound in Eq.\,(\ref{12}), the condition in Eq.~(\ref{lmass1}) for explaining the $R_{K^{(*)}}$ anomalies is fulfilled if $\cos(\phi_1+\phi_2) \approx 0.18$. We also note that for $M_{X_L} \approx 10 \ {\rm TeV} $ one could have $|\delta_i| \sim 0.02$, so the matrix $L^d$ in Eq.~(\ref{matricesL}) does not need to be highly tuned.
\vspace{1mm}

Finally, let us note that all loop-level constraints, including $K\!-\!\overline{K}$, $B\!-\!\overline{B}$, $B_s\!-\!\overline{B}_s$ mixing; radiative decays $\mu \to e\,\gamma$ (see App.~\ref{aapp1}), $\tau \to e\,\gamma$; anomalous magnetic and electric moments of leptons; $Z \to b\,\overline{b}$ and others \cite{Crivellin:2018yvo}, are satisfied due to the unitarity of  $L^d$  and a leptoquark mass in excess of $10 \ {\rm TeV}$.

\section{Collider phenomenology}

The aim of this limited phenomenological analysis is to simply demonstrate that the leptoquark $X_L$
in our model accounting for the flavor anomalies can be searched for at the next generation
collider. Focusing on the proposed $100 \ {\rm TeV}$ Future Circular Collider (FCC), we find that
one of the best signatures to look for is provided by the single leptoquark production process
\bea\label{ppjj} 
p \, p \rightarrow X_L\,j\, \mu^- \to j \,j\,\mu^+\mu^-\ .  
\eea 
In an in-depth
analysis one could also investigate final states involving other leptons, which for the case of
neutrinos would lead to missing energy signatures.  Pair production of $10 \ {\rm TeV}$ leptoquarks
is suppressed even at a $100 \ {\rm TeV}$ collider.\vspace{1mm}

To simulate the SM background and the leptoquark signal for the process (\ref{ppjj}) we used MadGraph 5 \cite{Alwall:2011uj} (version 2.6.3) with the default cuts apart from the lower cut on the transverse momentum of jets and leptons, which was set to $300 \ {\rm GeV}$. The leptoquark model file for MadGraph was implemented using FeynRules \cite{Alloul:2013bka} (version 2.3.32).
\vspace{1mm}

\begin{figure}[t!]
\includegraphics[width=0.95\linewidth]{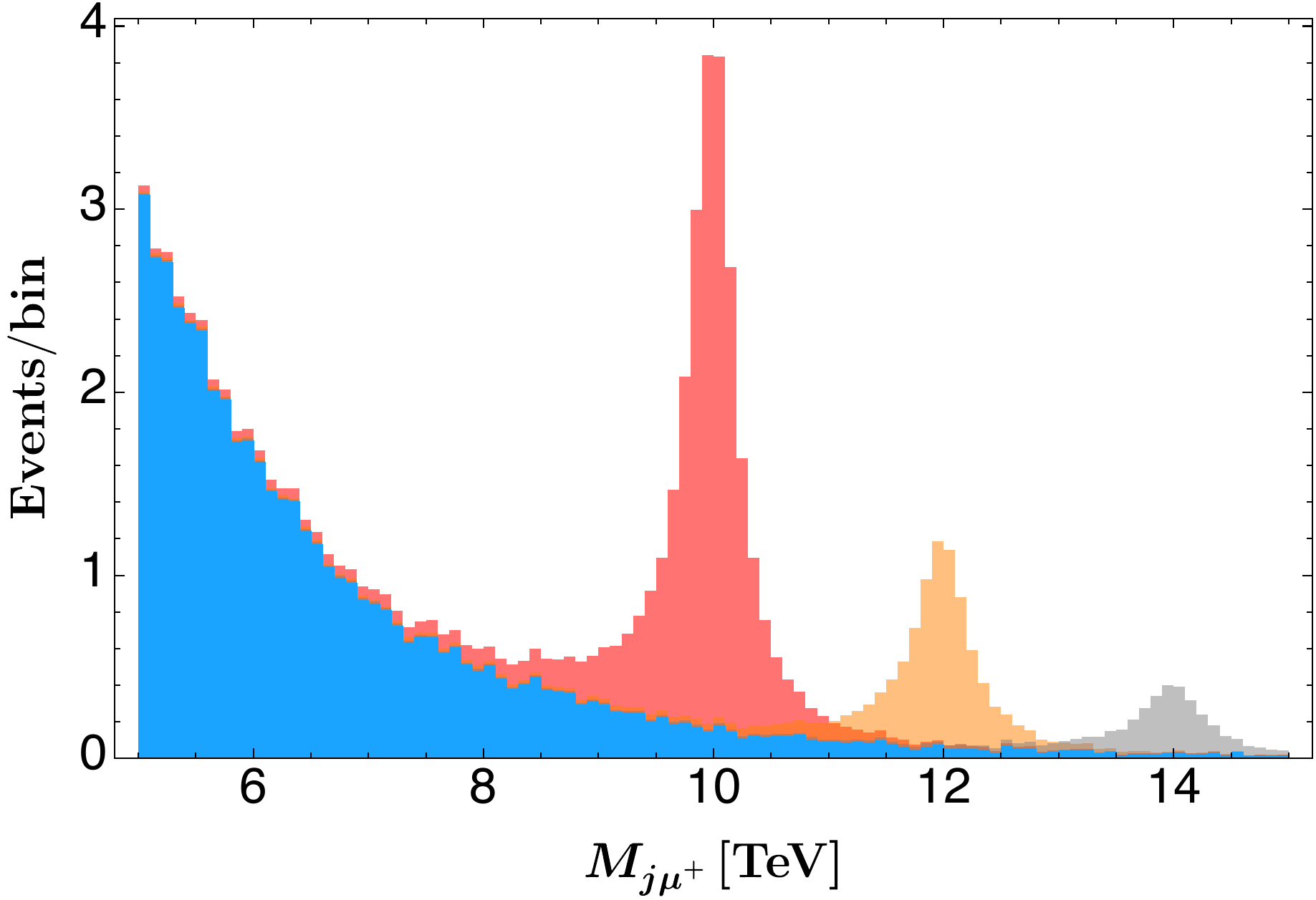}  
\caption{\small{Expected number of events in $250 \ {\rm fb}^{-1}$ of data collected by a $100 \ {\rm TeV}$ $p\hspace{0.3mm}p$  collider for the SM background $\,p \, p \to  \,j\,j \, \mu^+  \mu^-$ (blue) and the leptoquark signal $\,p \, p \to X_L\, j \, \mu^- \to \,j\,j \, \mu^+ \mu^-\,$  for masses $M_{X_L}= 10, 12, 14 \ {\rm TeV}$ (red, orange, gray) as a function of the invariant mass of the highest $p_T$ jet and $\mu^+$. The values of parameters discussed in Sec.~\ref{3} were used.\vspace{-1mm}}}
\label{fig1}
\end{figure}

Figure \ref{fig1} plots the number of background ($B$) and signal ($S$) events  for a leptoquark  mass $10$, $12$ and $14 \ {\rm TeV}$ expected within the first year of FCC running (estimated to be $250 \ {\rm fb}^{-1}$ of data \cite{Benedikt}) as a function of the invariant mass of the highest transverse momentum jet $j$ and $\mu^+$. Implementing the invariant mass cut 
$
|\,M_{j\mu^+} - M_{X_{L}}| < \Gamma_X \ ,
$
where $\Gamma_X$  is the width of the leptoquark, the significance of the signal, $S/\sqrt{B}$, is very high: 19 $\sigma$ for $M_{X_L} = 10 \ {\rm TeV}$, 6.7 $\sigma$ for $12 \ {\rm TeV}$ and 4.5 $\sigma$ for $14 \ {\rm TeV}$.
More sophisticated cuts may make the search  more efficient. A detailed analysis of the  $X_L$ vector leptoquark collider phenomenology is beyond the scope of this paper.\vspace{1mm}

Were the $B$ decay anomalies in $R_K$ and $R_{K^*}$ confirmed and established, inspection of
Eq.\,\eqref{lmass1} indicates that this model could be ruled out at a future 100~TeV high luminosity hadron
collider. Not only does the right-hand side of Eq.\,\eqref{lmass1} provide an upper bound on the
mass of the vector leptoquark, but  Eq.\,\eqref{gauge} shows that the strength of the coupling
constant $g_L$ is bounded from below,
and therefore the height of the resonant signal in Fig.~\ref{fig1} is bounded from below.\vspace{-1mm}

\section{Conclusions}
We have constructed a new model to account for the recently observed anomalies in $B$ meson decays set within the framework of Pati-Salam unification. The theory avoids all experimental bounds without introducing any vector-like fields mixing with the Standard Model fermions. This was achieved by suppressing the leptoquark right-handed interactions by associating them with a symmetry broken at a high scale, which eliminates the  most stringent constraints arising from the simultaneous presence of  left- and right-handed lepton flavor changing currents. In some regions of parameter space the mass of the leptoquark can be as low as $10 \ {\rm TeV}$ while remaining consistent with all experimental  data.
\vspace{1mm}

The tightest constraints on the model come from the experimental limits on rare kaon, $B$ meson and $\tau$ decays, as well as $\mu- e$ conversion. In the appendix we presented  general model-independent formulae for the various decay rates and listed the corresponding bounds. Those results can be used to read off the constraints on any model with one or more $(3,1)_{2/3}$ vector leptoquarks with arbitrary left- and right-handed interactions with Standard Model quarks and leptons.\vspace{1mm}

In our analysis we chose parameters to explain the $R_{K^{(*)}}$ flavor anomalies. As shown in \cite{Alonso:2015sja}, phenomenological models of the vector leptoquark $(3,1)_{2/3}$ can also account for $R_{D^{(*)}}$ anomalies \cite{Lees:2013uzd,Huschle:2015rga,Aaij:2015yra} (see \cite{Kumar:2018kmr} for an updated fit). The vector leptoquark $(3,1)_{2/3}$ in our model is too heavy to account also for the $R_{D^{(*)}}$ anomalies. Still, it has been shown \cite{Becirevic:2016yqi} that the scalar leptoquark $(3,2)_{1/6}$ might be a good candidate for that. This leptoquark appears in the scalar sector of our model and can be made sufficiently light. It would be interesting to investigate this in more detail.

Currently, there exist many models that account for the hints of lepton
  universality violation in $B$ meson decays. If these anomalies are
  established, new physics must emerge at a scale similar to that of
  the mass of the ``left-handed'' leptoquark in our model. We have
  demonstrated that simple kinematic cuts can isolate clearly
  observable signals with 250~fb$^{-1}$ of accumulated data at a
  100~TeV $p\hspace{0.3mm}p$ collider. Further analysis is badly required to determine
  whether such an apparatus could distinguish among the many proposed
  models.

\subsection*{Acknowledgments}
This research was supported in part by the DOE Grant No.~${\rm DE}$-${\rm SC0009919}$. 

\section*{Appendix}

\appendix

\section{\, ${\rm \bf SU}{\bf (4)}_{{\textbf{\textit{{L}}}}} \times {\rm \bf SU}{\bf (4)}_{{\textbf{\textit{{R}}}}}$\,  symmetry breaking}\label{aap3}
The scalar potential of the model is given by
\bea\label{potVp}
V = &-& \mu_{1}^2\, |\hat\Sigma_L|^2+ \lambda_{1} |\hat\Sigma_L|^4 -  \mu_{2}^2\, |\hat\Sigma_R|^2+ \lambda_{2} |\hat\Sigma_R|^4 - \mu_{3}^2\, |\hat\Sigma|^2\nn\\
&+&\lambda_3 (\hat\Sigma\hat\Sigma^\dagger)^2 + \lambda_{3}' |\hat\Sigma\hat\Sigma^\dagger|^2  - \mu_{4}^2\, |\hat{H}_d|^2+\lambda_4 \big(\hat{H}_d\hat{H}_d^\dagger\big)^2 \nn\\
&+& \lambda_{4}' |\hat{H}_d\hat{H}_d^\dagger|^2  - \mu_{5}^2\, |\hat{H}_u|^2+\lambda_5  \big(\hat{H}_u\hat{H}_u^\dagger\big)^2+ \lambda_{5}'  |\hat{H}_u\hat{H}_u^\dagger|^2  \nn\\[1pt]
&+& \lambda_{12} |\hat\Sigma_L|^2 |\hat\Sigma_R|^2 +  \lambda_{13} |\hat\Sigma_L|^2|\hat\Sigma|^2+ \lambda_{14} |\hat\Sigma_L|^2 |\hat{H}_d|^2 \nn\\[1pt]
&+&   \lambda_{15} |\hat\Sigma_L|^2 |\hat{H}_u|^2 + \lambda_{23}  |\hat\Sigma_R|^2 |\hat\Sigma|^2 +  \lambda_{24} |\hat\Sigma_R|^2 |\hat{H}_d|^2\nn\\[1pt]
&+& \lambda_{25}  |\hat\Sigma_R|^2 |\hat{H}_u|^2+  \lambda_{34}|\hat\Sigma|^2 |\hat{H}_d|^2 + \lambda_{35}  |\hat\Sigma|^2 |\hat{H}_u|^2\nn\\[1pt]
&+&  \lambda_{45}  |\hat{H}_d|^2 |\hat{H}_u|^2 +   \lambda_{13}'  |\hat\Sigma_{L}\hat\Sigma|^2 + \lambda_{14}'  |\hat\Sigma_L^{\dagger}\hat{H}_{d}|^2\nn\\[1pt]
&+&  \lambda_{15}' |\hat\Sigma_L^{\dagger}\hat{H}_{u}|^2+   \lambda_{23}'  |\hat\Sigma_{R}^\dagger\hat\Sigma|^2 + \lambda_{24}'   |\hat\Sigma_R\hat{H}_{d}|^2\nn\\[1pt]
&+&  \lambda_{25}' |\hat\Sigma_R\hat{H}_{u}|^2 +  \lambda_{34}'  |\hat\Sigma \hat{H}_{d}|^2 +  \lambda_{35}'|\hat\Sigma \hat{H}_{u}|^2\nn\\[1pt]
&+&  \lambda_{45}' |\hat{H}_d^\dagger\hat{H}_{u}|^2 +  \lambda_{34}'' \, {\rm Tr}(\hat\Sigma \hat{H}_{d} \hat{H}_{d}^\dagger \hat\Sigma^\dagger) \nn\\[1pt]
&+& \lambda_{35}'' \, {\rm Tr}(\hat\Sigma \hat{H}_{u} \hat{H}_{u}^\dagger \hat\Sigma^\dagger)  + \lambda_{45}'' \, {\rm Tr}(\hat{H}_{d}^\dagger \hat{H}_{u} \hat{H}_{u}^\dagger \hat{H}_{d}) \nn\\[1pt]
&+&  \big[  \lambda_{345}' (\hat{\Sigma}\hat{H}_{d})(\hat{\Sigma}\hat{H}_{u})  +  \lambda_{345}''  {\rm Tr}(\hat{\Sigma}\hat{H}_{d} \hat{\Sigma}\hat{H}_{u})\nn\\[1pt]
&+&   \lambda_{345}'''  {\rm Tr}(\hat{H}_{d} \hat{\Sigma}\hat{\Sigma}\hat{H}_{u})+\kappa\, \hat{\Sigma}_L\hat{\Sigma} \,\hat{\Sigma}_R^\dagger + {\rm h.c.} \big]\ ,
\eea
where we have adopted the notation:\vspace{1mm}\\
$
 |\hat\Sigma_L|^2\equiv (\hat\Sigma_L)_{\alpha_L} (\hat\Sigma_L^\dagger)^{\alpha_L}$, \  $|\hat\Sigma|^2 \equiv (\hat\Sigma)^{\alpha_L}_{\alpha_R} (\hat\Sigma^\dagger)^{\alpha_R}_{\alpha_L}$, \ $ (\hat\Sigma \hat{H}_d) \equiv  (\hat\Sigma)^{\alpha_L}_{\alpha_R}(\hat{H}_d)_{\alpha_L}^{\alpha_R}$, \ \ \ 
$|\hat\Sigma_{L}\hat\Sigma|^2 \equiv  (\hat\Sigma_{L})_{\alpha_L}(\hat\Sigma)^{\alpha_L}_{\alpha_R} (\hat\Sigma^\dagger)_{\beta_L}^{\alpha_R}(\hat\Sigma_{L}^\dagger)^{\beta_L}$,
$ |\hat\Sigma \hat{H}_d|^2 \equiv  (\hat\Sigma)^{\alpha_L}_{\alpha_R}(\hat{H}_d)_{\beta_L}^{\alpha_R} (\hat{H}_d^\dagger)^{\beta_L}_{\beta_R} (\hat\Sigma^\dagger)_{\alpha_L}^{\beta_R}$, \ ${\rm Tr}(\hat{\Sigma}\hat{H}_{d} \hat{\Sigma}\hat{H}_{u})\equiv (\hat{\Sigma})_{\alpha_R}^{\alpha_L}(\hat{H}_{d})^{\alpha_R}_{\beta_L} (\hat{\Sigma})^{\beta_L}_{\beta_R}(\hat{H}_{u})^{\beta_R}_{\alpha_L}$, \ etc.
\vspace{2mm}

Let us consider $\langle \hat{\Sigma}_L \rangle$,  $\langle \hat{\Sigma}_R \rangle$ and  $\langle \hat{\Sigma} \rangle$. Via a suitable ${\rm SU}(4)_L$ and  ${\rm SU}(4)_R$ transformation, it is possible to bring $\langle \hat{\Sigma}_L \rangle$ and  $\langle \hat{\Sigma}_R \rangle$ to the form
\bea
\langle \hat\Sigma_L\rangle = \frac{v_L}{\sqrt2}\!
\begin{pmatrix}
\,0\,\\
0\\
0\\
1
\end{pmatrix} , \ \ \ \ \langle \hat\Sigma_R\rangle =\frac{v_R}{\sqrt2}\!
\begin{pmatrix}
\,0\,\\
0\\
0\\
1
\end{pmatrix} \ ,
\eea
where $v_L$ and $v_R$ are real and positive.

\noindent
The remaining ${\rm SU}(3)$ invariance can be utilized to obtain
\bea
\langle \hat\Sigma\rangle = 
\begin{pmatrix}
a_1 & 0& 0& b_1\\
0 & a_2& 0& b_2\\
0 & 0& a_3& b_3\\
c_1 & c_2& c_3& d
\end{pmatrix} .
\eea
To argue that $\langle \hat{\Sigma} \rangle$ can be brought to the diagonal form as in Eq.\,(\ref{vevs}), it is sufficient to consider the potential terms $|\hat\Sigma|^2$, $(\hat\Sigma\hat\Sigma^\dagger)^2$, $|\hat\Sigma\hat\Sigma^\dagger|^2$, $|\hat\Sigma_{L}\hat\Sigma|^2$, $|\hat\Sigma_{R}^\dagger\hat\Sigma|^2$ and $\hat{\Sigma}_L\hat{\Sigma} \,\hat{\Sigma}_R^\dagger$. Since
\bea
&&\lambda_{13}'|\hat\Sigma_{L}\hat\Sigma|^2 = \tfrac12\lambda_{13}' v_L^2 (c_1^2+c_2^2+c_3^2 +d^2) \ ,\nn\\
&&\lambda_{23}'|\hat\Sigma_{R}^\dagger\hat\Sigma|^2 = \tfrac12\lambda_{23}' v_R^2 (b_1^2+b_2^2+b_3^2 +d^2) \ ,\\
&& - \mu_{3}^2\, |\hat\Sigma|^2+\lambda_3 (\hat\Sigma\hat\Sigma^\dagger)^2 = \lambda_3 \big(a_1^2+a_2^2+a_3^2 + b_1^2+b_2^2+b_3^2\nn\\[-2pt]
&&\hspace{29mm}+\ c_1^2+c_2^2+c_3^2+d^2 -v_\Sigma^2\big)^2 - \lambda_3 v_\Sigma^4\ ,\nn
\eea
the potential is minimized for $\langle \hat{\Sigma} \rangle  = {\rm diag}(a_1,a_2,a_3,d)$. 
In addition, the terms
\bea
&& \lambda_{3}' |\hat\Sigma\hat\Sigma^\dagger|^2 = \lambda_{3}' \big(a_1^4+a_2^4+a_3^4+d^4\big)\ ,\nn\\
&&\kappa \, \hat{\Sigma}_L\hat{\Sigma} \,\hat{\Sigma}_R^\dagger = \tfrac12\kappa\, v_L v_R \,d 
\eea
imply that the minimum occurs at $a_1=a_2=a_3$. Finally, we are
  free to choose $\kappa$ to be real and negative, which through an
  appropriate redefinition of $\hat{\Sigma}$ leads to real
    $d>0$; therefore
 \bea
\langle \hat\Sigma\rangle = \frac{v_\Sigma}{\sqrt2}
\begin{pmatrix}
\,1 & 0& 0& 0\,\\
\,0 & 1& 0& 0\,\\
\,0 & 0& 1& 0\,\\
\,0 & 0& 0& z\,
\end{pmatrix} 
\eea
with $z$ being real and positive. Note that only one of
  the parameters $\lambda'_{345}$, $\lambda''_{345}$ and
  $\lambda'''_{345}$ can be made real by a field redefinition. If
  any of the other two has a nonzero imaginary part, the scalar potential is $CP$-violating.  A rigorous minimization procedure is beyond the scope of this work.

\section{\,Scalar masses}\label{aap4}

To show that Eq.\,(\ref{expa}) can be satisfied, it is again sufficient to consider only a few terms in the scalar potential.  In terms of hard masses, the relevant part of the Lagrangian is
\bea
{\mathcal{L}_m} &\supset& \,M_d^2 \,|\hat{H}_d|^2 + M_u^2 \,  |\hat{H}_u|^2 \!- \lambda_{24}'  |\hat\Sigma_R\hat{H}_{d}|^2 \!-  \lambda_{25}' |\hat\Sigma_R\hat{H}_{u}|^2\nn\\
&&- \  \lambda_{34}'  |\hat\Sigma \hat{H}_{d}|^2 -  \lambda_{35}'   |\hat\Sigma \hat{H}_{u}|^2-   \lambda_{345}'  (\hat{\Sigma}\hat{H}_{d}) (\hat{\Sigma}\hat{H}_{u}) \ .\nn\\
\eea
This results in the masses for the color octets and triplets,
\bea
\begin{aligned}
&m_{O_1} = m_{T_1} = m_{T_2} \equiv M_d^2 \ ,\\
&m_{O_2} = m_{T_3} = m_{T_4} \equiv M_u^2 \ .
\end{aligned}
\eea
The  mass squared matrix for the  fields $S_{1,2,3,4}$ is
\bea\label{pmS}
&&{\mathcal{M}}_{S}^2 = \nn\\
&&\begin{pmatrix}
M_d^2+ a_d & 0 &\tfrac12\lambda_{345}'z^2v_\Sigma^2 & \tfrac{\sqrt3}{2}\lambda_{345}'z\,v_\Sigma^2\\[2pt]
0 & M_d^2 + b_d& \tfrac{\sqrt3}{2}\lambda_{345}'z\,v_\Sigma^2& \tfrac32\lambda_{345}'v_\Sigma^2\\[2pt]
\tfrac12\lambda_{345}'z^2v_\Sigma^2 & \tfrac{\sqrt3}{2}\lambda_{345}'z\,v_\Sigma^2& M_u^2+ a_u& 0\\[2pt]
\tfrac{\sqrt3}{2}\lambda_{345}'z\,v_\Sigma^2 & \tfrac32\lambda_{345}' v_\Sigma^2&0&M_u^2+b_u
\end{pmatrix} ,\nn\\
\eea
where 
\bea
&&a_d = \tfrac12\lambda_{24}'v_R^2 +\tfrac12\lambda_{34}'z^2 v_\Sigma^2 \ , \ \ \  b_d = \tfrac32\lambda_{34}' v_\Sigma^2 \ , \nn\\
&&a_u = \tfrac12\lambda_{25}'v_R^2 +\tfrac12\lambda_{35}'z^2 v_\Sigma^2 \ , \ \ \  b_u = \tfrac32\lambda_{35}' v_\Sigma^2 \ .
\eea
We have verified that there exists a class of solutions with only one linear combination of the four scalars being light. To reproduce the SM fermion masses while keeping the Yukawas  perturbative, it is sufficient to have the light mass eigenstate,  identified with the SM Higgs, given by
\bea\label{expakk}
H = - \,c_e\, S_1 - c_d\,S_2 + c_\nu\, S_3 + c_u\, S_4 \ ,
\eea
where $c_u \approx 1 \gg c_d \gg c_\nu$ and $ 1 \gg c_e \gg c_\nu$, with the ratio $c_d:c_e \approx m_b:m_\tau $.

\section{ \ Flavor constraints:  Model-independent analysis}\label{aapp1}

The general form of the  Lagrangian describing interactions of vector leptoquarks $(3,1)_{2/3}$ with fermions is given by
\bea\label{glag}
\mathcal{L}\  \ \supset \ \ && \sum_{\alpha} X^{(\alpha)}_{\mu} \Big[f^{Lu}_{ij(\alpha)}\,(\bar u^i\gamma^\mu P_L\nu^j) + f^{Ru}_{ij(\alpha)}\,(\bar u^i\gamma^\mu P_R\nu^j) \nonumber\\
&&+\  f^{Ld}_{ij(\alpha)}\,(\bar d^{\,i} \gamma^\mu P_L\, e^j) + f^{Rd}_{ij(\alpha)}\,(\bar d^{\,i} \gamma^\mu P_R\, e^j) \Big] \ ,
\eea
where the field $X^{(\alpha)}_\mu$ corresponds to a leptoquark with mass $M_\alpha$. The resulting contributions to rare processes are listed below, along with the most severe experimental bounds.

The numerical values for particle masses and lifetimes 
were adopted from PDG \cite{PDG}. The single-particle state normalization chosen is
\bea
\langle \,\vec{p}\, |\, \vec{p}\,' \rangle = 2E\,(2\pi)^3\delta^{(3)}(\vec{p}-\vec{p}\,') 
\eea
and the decay constant $f_{\mathcal{M}}$ for a meson consisting of quarks/antiquarks ${q}_1$, $q_2$ is defined via 
\bea
\begin{aligned}
\langle 0|\bar{q}_1 \gamma^5 q_2|\mathcal{M}(p)\rangle &=- i  f_{\mathcal{M}} \, \frac{m_{\mathcal{M}}^2}{m_{q_1}\!+m_{q_2}} \ , \\[3pt]
\langle 0|\bar{q}_1\gamma^\mu \gamma^5 q_2|\mathcal{M}(p)\rangle &= i  f_{\mathcal{M}} \, p^\mu \ .
\end{aligned}
\eea 

\vspace{1mm}
\noindent
The  following values for the meson decay constants were adopted from the PDG, 
\bea
&&f_{\pi^+} =130 \ {\rm MeV} \ , \ \ \  \ f_{K^0_L} = f_{K^+} = 156 \ {\rm MeV} \ , \\
&&f_{B^0} = 191 \ {\rm MeV} \ , \ \ \ \ f_{B^+} = 187 \ {\rm MeV} \ , \ \ \ \  f_{B^0_s} = 227 \ {\rm MeV} \ ,\nn
\eea
and were obtained by averaging the lattice results. 

\noindent
The  other decay constants needed in our analysis are
\bea
&&f_{\phi} \approx 238 \ {\rm MeV} \ , \  \ f_{\Upsilon(1S)} \approx 700 \ {\rm MeV} \ ,\nn\\
&&f_{\Upsilon(2S)} \approx 496 \ {\rm MeV} \ , \ \ f_{\Upsilon(3S)} \approx 430 \ {\rm MeV} \ ,
\eea
where $f_\phi$ was determined from the lattice \cite{Chakraborty:2017hry} and $f_{\Upsilon(nS)}$ $(n=1,2,3)$ were extracted from the experimental results for $\Upsilon(nS)\to \ell^- \ell^+$ (see \cite{Bhattacharya:2016mcc,Kumar:2018kmr}).

Below, we present constraints on a general model with $(3,1)_{2/3}$ vector leptoquarks described by the Lagrangian (\ref{glag}). The constraints arise from the following processes:

\vspace{5mm}
\begin{itemize}
\item[(1)] Neutral meson decays to two charged leptons
\vspace{-2mm}
\begin{itemize}
\item[(a)] Neutral kaon decays
\item[(b)] Neutral $B$ meson decays
\item[(c)] $\Upsilon$ decays
\vspace{0mm}
\end{itemize}
\item[(2)] Charged meson decays to a charged lepton and neutrino
\vspace{-6mm}
\begin{itemize}
\item[(a)] Charged pion decays
\item[(b)] Charged kaon decays
\vspace{0mm}
\end{itemize}
\item[(3)] Charged meson three-body decays to a meson and charged leptons
\vspace{-2mm}
\begin{itemize}
\item[(a)] Charged kaon decays
\item[(b)] Charged $B$ meson decays
\vspace{0mm}
\end{itemize}
\item[(4)] Tau  decays
\vspace{-1mm}
\item[(5)] Radiative charged lepton decay
\vspace{-1mm}
\item[(6)] $l_i^+ \!\to l_j^+$ \,conversion
\end{itemize}
The relevant formulae  are listed below. The numbering of the sections matches that in the table of contents above.\vspace{10mm}

\noindent
\centerline
{{(1)} \bf \ {\emph{Neutral meson decays to two charged leptons}}}\vspace{3mm}
The leptoquark contribution to the decay of a pseudoscalar meson $\mathcal{M}$ with mass $m_{\mathcal{M}}$ to two charged leptons, $l_i^+$ with mass $m_i$ and $l_j^-$ with mass  $m_j$, is given by
\bea\label{123}
&&\Gamma({\mathcal{M}} \rightarrow l_i^+l_j^-)_X= \ \frac{m_{\mathcal{M}}f^2_{\mathcal{M}}}{64\,\pi}\Bigg[\,A_{ij}\bigg(1-\frac{m_{i}^2+m^2_{j}}{m_{\mathcal{M}}^2}\bigg) \nonumber\\
&&+ \, B_{ij} \frac{4\,m_i m_j}{m_{\mathcal{M}}^2}\Bigg]\sqrt{\left[1\!-\!\frac{(m_{i}\!+\!m_{j})^2}{m_{\mathcal{M}}^2}\right]\left[1\!-\!\frac{(m_{i}\!-\!m_{j})^2}{m_{\mathcal{M}}^2}\right]}\ ,\nonumber\\
 \eea
 where
  \bea\label{mij5} 
  && \hspace{8mm}A_{ij} \equiv {\bigg|\sum_{\alpha}\frac{a_{ij(\alpha)}^{LR}}{M_\alpha^2}\bigg|^2}+ \, {\bigg|\sum_{\alpha}\frac{a_{ij(\alpha)}^{RL}}{M_\alpha^2}\bigg|^2} \ , \nn\\[2pt]
&& \hspace{8mm}B_{ij} \equiv \sum_{\alpha,\beta}\frac{{\rm Re}\Big[a_{ij(\alpha)}^{LR}a_{ij(\beta)}^{RL\,*}\Big]}{M_\alpha^2M_\beta^2}\ ,\nn\\[4pt]
 &&a_{ij(\alpha)}^{LR\,(K^0_L)} \equiv \frac{1}{\sqrt2}\Big[\,m_{i}\,f^{Rd}_{1i(\alpha)} f^{Rd\,*}_{2j(\alpha)} +m_{j}\,f^{Ld}_{1i(\alpha)} f^{Ld\,*}_{2j(\alpha)}\nn\\
 &&\hspace{20mm}+ \ \frac{2\,m_{{K^0}}^2Q}{m_{s}+m_{d}}f^{Ld}_{1i(\alpha)} f^{Rd\,*}_{2j(\alpha)}\Big] + (1 \leftrightarrow 2) \ , \nn\\[4pt]
  &&a_{ij(\alpha)}^{LR\,({B^0})} \equiv \ m_{i}\,f^{Rd\,*}_{1i(\alpha)} f^{Rd}_{3j(\alpha)} + m_{j}\,f^{Ld\,*}_{1i(\alpha)} f^{Ld}_{3j(\alpha)} \nn\\
  &&\hspace{20mm}+ \ \frac{2\,m_{B^0}^2Q}{m_{b}+m_{d}}f^{Ld\,*}_{1i(\alpha)} f^{Rd}_{3j(\alpha)} \  , \nn\\[4pt]
    &&a_{ij(\alpha)}^{LR\,({B^0_s})} \equiv \ m_{i}\,f^{Rd\,*}_{2i(\alpha)} f^{Rd}_{3j(\alpha)} + m_{j}\,f^{Ld\,*}_{2i(\alpha)} f^{Ld}_{3j(\alpha)} \nn\\
    &&\hspace{20mm}+ \ \frac{2\,m_{B^0_s}^2Q}{m_{b}+m_{s}}f^{Ld\,*}_{2i(\alpha)} f^{Rd}_{3j(\alpha)}  \ , \nn\\[4pt]
  &&a_{ij(\alpha)}^{RL} \equiv a_{ij(\alpha)}^{LR}\left(L\leftrightarrow R\right) .
  \eea
  
  In Eq.~(\ref{mij5}) the quark masses $m_q$ and the factor $Q$ depend on the energy scale, $m_q = m_q(\mu)$ and $Q = Q(\mu)$, with $Q(\mu)$ given by the formula
  \bea
 Q(\mu) = \left[\frac{\alpha^{(6)}(m_t)}{\alpha^{(6)}(M_{X_L})}\right]^{\frac47} \left[\frac{\alpha^{(5)}(m_b)}{\alpha^{(5)}(m_t)}\right]^{\frac{12}{23}}  \left[\frac{\alpha^{(4)}(\mu)}{\alpha^{(4)}(m_b)}\right]^{\frac{12}{25}}  \!\!, \ \ \ \ \ \ \ \ \ 
 \eea
applicable for $m_b > \mu > m_c$. The  coupling constant $\alpha$  is calculated from
 \bea
 \alpha^{(N_f)}(\mu, \Lambda) = \frac{4\,\pi}{(11-2N_f/3)\log(\mu^2/\Lambda^2)} \ ,
 \eea
where $N_f$ is the number of quark flavors at a given scale, by matching 
\bea
 &&\alpha^{(6)}(m_t) \equiv  \alpha^{(6)}(m_t, \Lambda_6) =  \alpha^{(5)}(m_t, \Lambda_5) \equiv  \alpha^{(5)}(m_t), \ \ \ \ \ \ \ \  \nn\\
 && \alpha^{(5)}(m_b) \equiv  \alpha^{(5)}(m_b, \Lambda_5) =  \alpha^{(4)}(m_b, \Lambda_4) \equiv  \alpha^{(4)}(m_b).\nn\\
\eea
 The ratio $Q(\mu)/m_q(\mu)$ is a renormalization group invariant.
Adopting the PDG values for the quark masses at  $\mu = 2 \ {\rm GeV}$  and for the strong coupling constant at $\mu = M_Z$ \cite{PDG}, the value  of $Q$ depends only on the leptoquark mass scale through
\bea
Q(2 \ {\rm GeV}) = \frac{0.45}{[{\alpha^{(6)}(M_{X_L})}]^{4/7} }\ .
\eea

As evident from Eq.~(\ref{mij5}), the constraints on the leptoquark contribution to the branching fraction of kaon and $B$ meson decays are much weaker when the leptoquarks have only LH or only RH interactions with SM fermions, as opposed to models with both LH and RH interactions. The bounds on the branching fraction are milder by a factor of
\bea
\sqrt{2\,m_{\mathcal{M}}^2Q / {(m_l\,m_q)}} \ ,\nn
\eea 
which is reflected by the much weaker constraints  on the leptoquark mass in our model compared to generic leptoquark models (see App.~\ref{aapp2}).

 For the majority of decays considered here only the upper bound on the rate was experimentally established.  However, in the four cases: $K^0_L \rightarrow e^+e^-$, $K^0_L \rightarrow \mu^+\mu^-$, $B^0 \rightarrow \mu^+\mu^-$ and $B_s^0 \rightarrow \mu^+\mu^-$ nonzero rates have been measured. For those particular decays not only the pure leptoquark contribution is relevant, but also the interference effects with the SM short-distance (SD) contribution. This can be taken into account  by making the following substitution in the expressions for $A_{ij}$ and $B_{ij}$ in Eq.~(\ref{mij5}):
\bea\label{subs}
\sum_{\alpha}\frac{a_{ij(\alpha)}^{LR}}{M_\alpha^2}  \ \longrightarrow \ \sqrt{\frac{64\,\pi\,\Gamma({\mathcal{M}} \!\rightarrow\! l_i^+l_j^-)^{\rm SD}_{\rm SM}}{{m_{\mathcal{M}}f^2_{\mathcal{M}}}}}\,\delta_{ij}  \pm \sum_{\alpha}\frac{a_{ij(\alpha)}^{LR}}{M_\alpha^2} \ ,\nn\\
\eea
where the $+/-$ depends on the decay considered and corresponds to the SM short-distance amplitude for ${\mathcal{M}} \rightarrow l_i^+l_j^-$ being negative/positive. The leptoquark-induced contribution is then obtained by subtracting off the pure SM part.\\
 
\noindent
\centerline
{{(a)} \bf \ {\emph{Neutral kaon decays}}}\vspace{3mm}
The decays $K_L^0 \rightarrow e^\pm\mu^\mp$ are absent in the SM and the constraint on the leptoquark mass is derived directly from the experimental bound  on the branching fraction, 
$
{\rm Br}_X \lesssim \Delta{\rm Br}.
$
The rates for $K_L^0 \rightarrow e^+e^-\!, \, \mu^+\mu^-$ were measured \cite{Ambrose:1998cc,PDG}. They are dominated by long-distance SM effects  \cite{Valencia:1997xe,Isidori:2003ts}. 
For $K_L^0 \rightarrow e^+e^-$ the experimental branching fraction $ \left(8.7^{+5.7}_{-4.1}\right) \times 10^{-12}$  \cite{Ambrose:1998cc} agrees well with the SM long-distance estimate of $(9.0\pm 0.5) \times 10^{-12}$ \cite{Valencia:1997xe}. In that case we use the experimental uncertainty for the measured branching fraction as the upper bound for the leptoquark contribution.
For $K_L^0 \rightarrow \mu^+\mu^-$ the measured branching fraction is $(6.84 \pm 0.11) \times 10^{-9}$ \cite{Ambrose:2000gj}, but it was shown  that the short-distance SM contribution  is only $0.9 \times 10^{-9}$ \cite{Valencia:1997xe}, whereas the upper bound on the total short-distance contribution is $2.5 \times 10^{-9}$ \cite{Isidori:2003ts}. 

The constraints below reflect the most conservative bound on the leptoquark mass obtained using Eq.~(\ref{123}) (the branching fractions were left in explicitly for easier use of the formulae given  future experimental improvements):
\bea\label{limits1}
 &&  \ \ \  \frac{A_{11}^{({K_L^0})}}{m_{K^0}^2} \ \lesssim \ \left[\,\frac{{\rm Br}(K_L^0\rightarrow e^+e^-)}{5.7 \times 10^{-12}}\right] (672 \ \,{\rm TeV})^{-4} \ ,\hspace{6mm}\text{\cite{Ambrose:1998cc}}\nn\\
 &&\\[4pt]
   && \ \ \ \frac{A_{12}^{(K_L^0)}+A_{21}^{(K_L^0)}}{m_{K^0}^2} \ \lesssim \ \left[\,\frac{{\rm Br}(K_L^0 \rightarrow e^\pm \mu^\mp)}{4.7 \times 10^{-12} }\right]\hspace{14.3mm} \text{\cite{Ambrose:1998us}} \nn\\
 && \hspace{33mm} \times \ (689\ \,{\rm TeV})^{-4} \ , \\[10pt]
  && \ \ \ \frac{{A'}_{\!22}^{({K_L^0})}+ 0.2\,{B'}_{\!\!22}^{({K_L^0})}}{m_{K^0}^2} \ \lesssim \ \left[\,\frac{{\rm Br}(K_L^0\rightarrow \mu^+\mu^-)}{2.5 \times 10^{-9}}\right] \hspace{6.5mm} \text{\cite{Isidori:2003ts}}\nn\\
  && \hspace{40.5mm}\times \  (140 \ \,{\rm TeV})^{-4} \ ,
 \eea
where ${A'}_{\!22}^{(K_L^0)}$ is given by  ${A}_{22}^{(K^0_L)}$ with the substitution (\ref{subs}) with ${\Gamma(K_L^0\rightarrow \mu^+\mu^-)^{\rm SD}_{\rm SM}} = 0.9 \times 10^{-9}$; similarly for ${B'}_{\!\!22}^{(K_L^0)}$.

\centerline
{{(b)} \bf \ {\emph{Neutral B meson decays}}}\vspace{3mm}

For most of the $B^0$ and $B^0_s$ decays only the limit on the branching fraction is determined; therefore the bounds on leptoquark parameters are derived using
$
{\rm Br}_X \lesssim \Delta{\rm Br}.
$
In the case of $B^0 \rightarrow \mu^+\mu^-$ and $B_s^0 \rightarrow \mu^+\mu^-$  the  branching fractions were actually measured, $ {\rm Br}({B^0} \rightarrow \mu^+\mu^-) = \left(1.6^{+1.6}_{-1.4}\right) \times 10^{-10}$  \cite{PDG} and $ {\rm Br}({B^0_s} \rightarrow \mu^+\mu^-) = \left(3.0\pm0.6^{+0.3}_{-0.2}\right) \times 10^{-9}$ \cite{Aaij:2017vad}, and they are dominated by short-distance SM effects. We arrive at the following 
set of constraints:

\bea
 &&  \ \ \  \frac{A_{11}^{({B^0})}}{m_{B^0}^2} \ \lesssim \ \left[\,\frac{{\rm Br}({B^0} \rightarrow e^+e^-)}{8.3 \times 10^{-8}}\right] (29.4 \ \,{\rm TeV})^{-4}\ ,\hspace{6.4mm}\text{\cite{Aaltonen:2009vr}}\nn\\
 &&\\[4pt]
  &&  \ \ \ \frac{A_{12}^{(B^0)}+A_{21}^{(B^0)}}{m_{B^0}^2} \ \lesssim \ \left[\,\frac{{\rm Br}(B^0 \rightarrow e^\pm\mu^\mp)}{1.0 \times 10^{-9}}\right]\hspace{16.2mm}  \text{\cite{Aaij:2017cza}} \nn\\
 && \hspace{32mm} \times \ (88.6 \ \,{\rm TeV})^{-4} \ , \\[7pt]
  &&  \ \ \  \frac{{A'}_{\!22}^{({B^0})}}{m_{B^0}^2} \ \lesssim \ \left[\,\frac{{\rm Br}({B^0} \rightarrow \mu^+\mu^-)}{1.6\times 10^{-10}}\right] (140 \ \,{\rm TeV})^{-4}\ ,\hspace{5.4mm}\text{\cite{PDG}}\nn\\
 &&\\[4pt]
   &&  \ \ \ \frac{A_{13}^{(B^0)}+A_{31}^{(B^0)}}{m_{B^0}^2} \ \lesssim \ \left[\,\frac{{\rm Br}(B^0 \rightarrow e^\pm\tau^\mp)}{2.8 \times 10^{-5} }\right]\hspace{16.2mm}  \text{\cite{Aubert:2008cu}} \nn\\
 && \hspace{32mm} \times \ (6.4 \ \,{\rm TeV})^{-4} \ , \\[7pt]
    &&  \ \ \ \frac{A_{23}^{(B^0)}+A_{32}^{(B^0)}}{m_{B^0}^2} \ \lesssim \ \left[\,\frac{{\rm Br}(B^0 \rightarrow \mu^\pm\tau^\mp)}{2.2 \times 10^{-5} }\right]\hspace{15.7mm}  \text{\cite{Aubert:2008cu}} \nn\\
 && \hspace{32mm} \times \ (6.8 \ \,{\rm TeV})^{-4} \ , \\[7pt]
     && \ \ \ \frac{A_{33}^{(B^0)}+0.59\,B_{33}^{(B^0)}}{m_{B^0}^2} \ \lesssim \ \left[\,\frac{{\rm Br}(B^0 \rightarrow \tau^+\tau^-)}{2.1\times 10^{-3}}\right]\hspace{8.9mm} \text{\cite{Aaij:2017xqt}} \nn\\
 && \hspace{38.7mm} \times \ (2.0 \ \,{\rm TeV})^{-4} \ , \\[7pt]
 && \ \ \  \frac{A_{11}^{({B^0_s})}}{m_{B^0_s}^2} \ \lesssim \ \left[\,\frac{{\rm Br}({B^0_s} \rightarrow e^+e^-)}{2.8 \times 10^{-7}}\right] (23.9 \ \,{\rm TeV})^{-4}\ ,\hspace{6.4mm}\text{\cite{Aaltonen:2009vr}} \nn\\
 &&\\[4pt]
  && \ \ \ \frac{A_{12}^{(B^0_s)}+A_{21}^{(B^0_s)}}{m_{B^0_s}^2} \ \lesssim \ \left[\,\frac{{\rm Br}(B^0_s \rightarrow e^\pm\mu^\mp)}{5.4 \times 10^{-9}}\right]\hspace{16.1mm}  \text{\cite{Aaij:2017cza}}  \nn\\
 && \hspace{32mm} \times \ (64.1 \ \,{\rm TeV})^{-4} \ , \\[7pt]
   &&  \ \ \  \frac{{A'}_{\!22}^{({B^0_s})}}{m_{B^0_s}^2} \ \lesssim \ \left[\,\frac{{\rm Br}({B^0_s} \rightarrow \mu^+\mu^-)}{0.7\times 10^{-9}}\right] (107 \ \,{\rm TeV})^{-4}\ ,\hspace{5.5mm}\text{\cite{Aaij:2017vad}}\nn\\
 &&\\[4pt]
      &&  \ \ \  \frac{A_{33}^{(B^0_s)}+0.56\,B_{33}^{(B^0_s)}}{m_{B^0_s}^2} \ \lesssim \ \left[\,\frac{{\rm Br}(B^0_s \rightarrow \tau^+\tau^-)}{6.8\times 10^{-3}}\right]\hspace{9mm} \text{\cite{Aaij:2017xqt}} \nn\\
 && \hspace{39mm} \times \ (1.7 \ \,{\rm TeV})^{-4} \ ,
 \eea
where ${A'}_{\!22}^{(B^0)}$ is given by  ${A}_{22}^{(B^0)}$ with the substitution (\ref{subs}) with ${\Gamma(B^0\rightarrow \mu^+\mu^-)^{\rm SD}_{\rm SM}} = 1.6 \times 10^{-10}$, and similarly for ${A'}_{\!22}^{(B^0_s)}$ with ${\Gamma(B^0_s\rightarrow \mu^+\mu^-)^{\rm SD}_{\rm SM}} = 3.0 \times 10^{-9}$.
We listed the constraint on the leptoquark contribution to ${\rm Br}(B_s^0 \rightarrow \mu^+\mu^-)$ for completeness, but this branching fraction is actually determined by the fit that yields $\Delta C_9$ and $\Delta C_{10}$  in Eq.~(\ref{c9}).\\

\vspace{1mm}
\centerline
{{(c)} \bf \ {\emph{$\boldsymbol\Upsilon$ decays}}}\vspace{3mm}

This set of constraints arises from the vector meson decays $\Upsilon(nS)\to e^\pm \tau^\mp$  and $\Upsilon(nS)\to \mu^\pm \tau^\mp$.  Neglecting the electron and muon mass, the corresponding branching fraction is given by Eq.\,(\ref{123}) (only the term with $A_{ij}$ is nonzero), where
\bea
      &&a_{ij(\alpha)}^{LR\,(\Upsilon(nS))} \equiv\ \sqrt{\frac{8}{3}\bigg(1+\frac{m_\tau^2}{2m_{\Upsilon(nS)}^2}\bigg)}\ \bigg(m_{i}\,f^{Rd\,*}_{3i(\alpha)} f^{Rd}_{3j(\alpha)} \nn\\
      && \hspace{20mm}+ \ m_{j}\,f^{Ld\,*}_{3i(\alpha)} f^{Ld}_{3j(\alpha)}+  \frac{m_{\Upsilon(nS)}^2Q}{m_{b}}f^{Ld\,*}_{3i(\alpha)} f^{Rd}_{3j(\alpha)}\bigg) \, ,\nn\\[4pt]
        &&a_{ij(\alpha)}^{RL\,(\Upsilon(nS))} \equiv a_{ij(\alpha)}^{LR\,(\Upsilon(nS))}\left(L\leftrightarrow R\right) .
\eea
\vspace{0mm}

\noindent
Of all $\Upsilon(nS)\to \ell^\pm \tau^\mp$ decays, the $\Upsilon(3S)\to \mu^\pm \tau^\mp$ gives the tightest constraints. From the corresponding experimental bounds we have
\bea
   &&  \ \ \ \frac{A_{23}^{(\Upsilon(3S))}+A_{32}^{(\Upsilon(3S))}}{m_{\Upsilon(3S)}^2} \lesssim  \left[\,\frac{{\rm Br}(\Upsilon(3S)\to \mu^\pm \tau^\mp)}{3.1 \times 10^{-6}}\right]\hspace{4mm}  \text{\cite{Aaij:2017cza}} \nn\\
 && \hspace{38mm} \times \ (0.5 \ \,{\rm TeV})^{-4} \ , 
\eea
which, however, is still much weaker than all other constraints considered here.\vspace{6mm}

\noindent
\centerline
{{(2)} \bf \ {\emph{Charged meson decays to a}}} \\ 
\centerline{\bf {\emph{ \ \ \ \ \ charged lepton and a  neutrino}}}\vspace{3mm}

Decays of mesons to a charged lepton and a neutrino exist in the SM. The leading order leptoquark contribution comes from interference effects. The theoretical uncertainty in the SM calculation is reduced by taking ratios of decay rates, 
\bea\label{1234}
\frac{\Gamma({\mathcal{M}} \rightarrow l_i^+ \nu)}{\Gamma({\mathcal{M}} \rightarrow l_{i'}^+ \nu)}&=&  \frac{\Gamma({\mathcal{M}} \rightarrow l_i^+ \nu)}{\Gamma({\mathcal{M}} \rightarrow l_{i'}^+ \nu)}\bigg|_{\rm SM}
\left(1+\frac{D_{i'}-D_{i}}{\sqrt2 \, G_F}\right), \nn\\
  \eea
  
  \vspace{-3mm}
  \noindent
 where
 
 \vspace{-6mm}
  \bea\label{mij14}
 && D_{i} \,\equiv\, \sum_{\alpha ; \,j} \frac{1}{M_\alpha^2}\,{{\rm Re}\Big[d_{ij(\alpha)}^{LR}+d_{ij(\alpha)}^{RL}\Big]} \ .
 \eea
 For the case of Dirac neutrinos,
 \bea
   &&d_{ij(\alpha)}^{LR\,(\pi^+)}\, \equiv \ \frac{U_{ij}}{V_{ud}}\nn\\[-2pt]
   &&\hspace{10mm}\times \left[f^{Rd}_{1i(\alpha)} f^{Ru\,*}_{1j(\alpha)} + \frac{2\,m_{\pi^+}^2Q}{m_i(m_{d}+m_{u})}f^{Ld}_{1i(\alpha)} f^{Ru\,*}_{1j(\alpha)}\right]  , \nn\\
 &&d_{ij(\alpha)}^{LR\,(K^+)} \,\equiv \ \frac{U_{ij}}{V_{us}}\nn\\[-2pt]
 &&\hspace{10mm}\times  \left[f^{Rd}_{1i(\alpha)} f^{Ru\,*}_{2j(\alpha)} + \frac{2\,m_{{K^+}}^2Q}{m_i(m_{s}+m_{u})}f^{Ld}_{1i(\alpha)} f^{Ru\,*}_{2j(\alpha)}\right] , \nn\\[4pt]
  &&d_{ij(\alpha)}^{RL} \equiv d_{ij(\alpha)}^{LR}\left(L\leftrightarrow R\right)  ,
 \eea

 \noindent
 whereas for Majorana neutrinos the only nonzero terms are
 \bea
d_{ij(\alpha)}^{RL\,(\pi^+)}\, \!\!\equiv\!\! \ \frac{U_{ij}}{V_{ud}} f^{Ld}_{1i(\alpha)} f^{Lu\,*}_{1j(\alpha)} \  , \ \ d_{ij(\alpha)}^{RL\,(K^+)}\, \!\!\equiv\!\! \ \frac{U_{ij}}{V_{sd}} f^{Ld}_{1i(\alpha)} f^{Lu\,*}_{2j(\alpha)}  \ .\nn\\
   \eea
 The tightest bounds of this type originate from measurements of the branching fraction ratios:
 \bea
R(\pi^+) \equiv  \frac{\Gamma(\pi^+ \rightarrow e^+ \nu)}{\Gamma(\pi^+ \rightarrow \mu^+ \nu) } \ , \ \  \ 
R(K^+) \equiv  \frac{\Gamma(K^+ \rightarrow e^+ \nu)}{\Gamma(K^+ \rightarrow \mu^+ \nu) } \ . \nn
 \eea

\vspace{3mm}
\centerline
{{(a)} \bf \ {\emph{Charged pion decays}}}\vspace{3mm}

\noindent
The experimental measurement  and the SM prediction yield
\bea
R(\pi^+) \ &=& (1.2327\pm0.0023)\times 10^{-4}  \ \ \ \  \text{\cite{PDG}}  \ , \nn\\
 R(\pi^+)_{\rm SM} &=& (1.2352\pm0.0001)\times 10^{-4}  \ \ \ \ \text{\cite{Cirigliano:2007ga}}  \ , \nn
 \eea
 which, given Eq.~(\ref{1234}), leads to 
\bea
\big|D_{1}^{(\pi^+)} - D_{2}^{(\pi^+)}\big| \ \lesssim \  (3.9 \ \,{\rm TeV})^{-2} \ . \ \ \ \ \ \ 
\eea

\vspace{3mm}
\centerline
{{(b)} \bf \ {\emph{Charged kaon decays}}}\vspace{3mm}

\noindent
In this case,
\bea
R(K^+) \ &=& (2.493\pm0.031)\times 10^{-5} \ \ \ \ \text{\cite{Ambrosino:2009aa}}\ , \nn\\
 R(K^+)_{\rm SM} &=&  (2.477\pm0.001)\times 10^{-5}  \ \ \ \ \text{\cite{Cirigliano:2007ga}} \ , \nn
 \eea
 which results in
 \bea
 \big|D_{1}^{(K^+)} - D_{2}^{(K^+)}\big| \lesssim \ (3.1 \ \,{\rm TeV})^{-2} \ . \ \ \ 
 \eea
\vspace{4mm}

\noindent
\centerline
{{(3)} \bf \ {\emph{Charged meson three-body decays}}} \\ 
\centerline{\bf {\emph{ \ \ \ \ \ to a meson and charged leptons}}}\vspace{3mm}

When the leptoquark has both LH and RH interactions with SM fermions, the three-body meson decays  are less restrictive than the two-body decays. However, in the case of our model, with predominantly LH interactions, the bounds arising from $B^+\to K^+e^\pm\mu^\mp$ impose the most severe constraints on the leptoquark mass.
The corresponding decay rate is expressed in terms of the form factors $f_+(q^2)$ and $f_0(q^2)$  defined via
\bea
&&\langle\mathcal{M}'(p') | \bar{q}_1\gamma^\mu q_2 |\overline{\mathcal{M}}(p)\rangle\nn\\
&& \hspace{0mm}= f_+(q^2)\left[ \, p^\mu+{p'}^\mu-\frac{\Delta M^2}{q^2}\, q^\mu\right]+ f_0(q^2) \frac{\Delta M^2}{q^2}q^\mu \ ,\nn\\[3pt]
&&\langle\mathcal{M}'(p') | \bar{q}_1 q_2 |\overline{\mathcal{M}}(p)\rangle = f_0(q^2)\frac{\Delta M^2}{m_{q_1} - m_{q_2}}\ ,
\eea
where  the four-momentum transfer $q=p'-p$ and the meson squared mass difference 
$
\Delta M^2 = m_{\mathcal{M}}^2- m_{\mathcal{M}'}^2$.
The contribution to the decay rate mediated by leptoquarks is
\bea
&& \Gamma(\mathcal{M}\rightarrow \mathcal{M}' \,l_i^+l_j^- )_X \ = \ \frac{1}{2048\,\pi^3}\frac{1}{m_\mathcal{M}^3} \nn\\
&&\times  \int_{(m_i+m_j)^2}^{(m_\mathcal{M}-m_{\mathcal{M}'})^2}\!\frac{dq^2}{q^4}\sqrt{\lambda(q^2, m_{\mathcal{M}}^2, m_{\mathcal{M}'}^2)\ \lambda(q^2, m_{i}^2, m_{j}^2) \ }\nonumber\\
&&\times \  \Bigg\{ \bigg[\,\frac13\,N^+_{ij}\,\bigg(2\,q^2-m_i^2-m_j^2- \frac{(m_i^2-m_j^2)^2}{q^2}\,\bigg)\nn\\[0pt]
&& \ \ \ \ \ \ \ \ \ + \ 2\,N^-_{ij}\,m_i\,m_j\bigg] \  \lambda(q^2, m_{\mathcal{M}}^2, m_{\mathcal{M}'}^2)\ |f_{+}(q^2)|^2\nonumber\\[3pt]
&&+ \ \bigg[\,N^+_{ij}\,\bigg(m_i^2+m_j^2-\frac{{(m_i^2-m_j^2)^2}}{q^2}\bigg) \,-\, 2\,N^-_{ij}\,m_i\,m_j \nonumber\\[3pt]
&& \ \ \ \ \ \ + \ 4\,P^+_{ij} \,q^2{(q^2-m_i^2-m_j^2)} - 8\,P^-_{ij} \,q^2\,m_i \,m_j\nn\\[5pt]
&& \ \ \ \ \ \ - \ 4\,\big(R^+_{ij}+R^-_{ij}\big)\,m_i \,\big(q^2-m_i^2+m_j^2\big)\nn\\[0pt]
&& \ \ \ \ \ \ + \ 4\,\big(R^+_{ij}-R^-_{ij}\big)\,m_j\,\big(q^2+m_i^2-m_j^2\big)\,\bigg] \nn\\[-4pt]
&& \ \ \ \ \times \  \left(m_{\mathcal{M}}^2- {m^2_{\mathcal{M}'}}\right)^{\!2} \,|f_{0}(q^2)|^2\Bigg\} \ ,
\eea

\vspace{-2mm}
\noindent
where
 \bea\label{mij2}
 && \, \lambda(x,y,z) \,\equiv\, \left(x-y-z\right)^2 -4\,y \,z \ , \nn\\[4pt]
 && N^\pm_{ij}\equiv  {\bigg|\sum_{\alpha}\frac{n_{ij(\alpha)}^{LL}+\, n_{ij(\alpha)}^{RR}}{M_\alpha^2}\bigg|^2} \pm \, {\bigg|\sum_{\alpha}\frac{n_{ij(\alpha)}^{LL}-\, n_{ij(\alpha)}^{RR}}{M_\alpha^2}\bigg|^2}
\ , \nn\\[5pt]
 && P^\pm_{ij} \equiv  {\bigg|\sum_{\alpha}\frac{p_{ij(\alpha)}^{LR}+\,p_{ij(\alpha)}^{RL}}{M_\alpha^2}\bigg|^2} \pm\,  {\bigg|\sum_{\alpha}\frac{p_{ij(\alpha)}^{LR}-\,p_{ij(\alpha)}^{RL}}{M_\alpha^2}\bigg|^2}\ , \nn\\[5pt]
&&R^{\pm}_{ij} \equiv \sum_{\alpha,\beta}\frac{{\rm Re}\Big[\Big(n_{ij(\alpha)}^{LL}\pm\, n_{ij(\alpha)}^{RR}\Big)\Big(p_{ij(\beta)}^{LR\,*}\pm\,p_{ij(\beta)}^{RL\,*}\Big)\Big]}{M_\alpha^2\,M_\beta^2} \ ,\nn
\eea
\bea
  && {n_{ij(\alpha)}^{LL\,(K^+\!,\pi^+)}} \!=  f^{Ld}_{1i(\alpha)} f^{Ld\,*}_{2j(\alpha)} \ , \ \ \ {n_{ij(\alpha)}^{LL(\, B^+\!,\pi^+)}} \!=  f^{Ld}_{1i(\alpha)} f^{Ld\,*}_{3j(\alpha)}  \ ,\nn\\[5pt]
  &&{n_{ij(\alpha)}^{LL\,(B^+\!,K^+)}} \!=  f^{Ld}_{2i(\alpha)} f^{Ld\,*}_{3j(\alpha)} \ , \ \ \ {n_{ij(\alpha)}^{RR}} = {n_{ij(\alpha)}^{LL}}\left(L \leftrightarrow R \right) \, ,\nn\\[5pt]
    && {p_{ij(\alpha)}^{LR\,(K^+\!\!,\pi^+\!)}} \!= \! \frac{f^{Ld}_{1i(\alpha)} f^{Rd\,*}_{2j(\alpha)}Q}{m_s-m_d}  , \ \, {p_{ij(\alpha)}^{LR\,(B^+\!\!,\pi^+\!)}} \!= \! \frac{f^{Ld}_{1i(\alpha)} f^{Rd\,*}_{3j(\alpha)}Q}{m_b-m_d}    ,\nn\\[5pt]
&&   {p_{ij(\alpha)}^{LR\,(B^+\!\!,K^+)}} \!=  \frac{f^{Ld}_{2i(\alpha)} f^{Rd\,*}_{3j(\alpha)}Q}{m_b-m_s} \, , \ \ \ {p_{ij(\alpha)}^{RL}} = {p_{ij(\alpha)}^{LR}}\left(L \leftrightarrow R \right)  \ .\nn\\
 \eea
 \noindent
 
 The form factors $f_{+}(q^2)$ and $f_0(q^2)$ are calculated using lattice methods. For the $K\to\pi$ form factor we use the linear fit given in \cite{Aoki:2017spo}. For the $B\to\pi$ and  $B\to K$ form factors we adopt the results of \cite{Flynn:2015mha} and \cite{Bouchard:2013pna}, respectively, where the interpolating functions were obtained using the Bourrely-Caprini-Lellouch parameterization \cite{Bourrely:2008za}.
\\ \\
The $K \to \pi$ \,form factor is \cite{Aoki:2017spo}
\bea
f_{\{+,0\}}(q^2)_{K\pi} = f_{+}(0)_{K \pi}\left[1+{\lambda'}_{\!\{+,0\}}\frac{q^2}{m_{\pi^+}^{2}}\right] \ ,\ \ \ \ \ \ \ \ \ \ 
\eea
with $f_+(0)_{K \pi} = 0.9636$, \ ${\lambda'}_{\!+} = 0.0308$ \,and \ ${\lambda'}_{\!0} = 0.0198$ .  \\ \\
The $B \to \pi$ \,and\, $B \to K$\, form factors are given by
\bea
&&f_{+}(q^2) = \frac{1}{P_+(q^2)} \sum_{n=0}^{N_+-1} b_+^{(n)} \left[z^n-(-1)^{n-N_+} \frac{n}{N_+} \,z^{N_+}\right] , \ \nn\\
&&f_{0}(q^2) = \sum_{n=0}^{N_0} b_0^{(n)} z^n \ , \ \ \ \ \ \ \ \ \ \ 
\eea
where 

\vspace{-8mm}
\bea
z \equiv z(q^2) = \frac{\sqrt{t_+-q^2} -\sqrt{t_+-t_0}}{\sqrt{t_+-q^2} +\sqrt{t_+-t_0}} \ . \ \ \ 
\eea

\noindent
In the  \,$B \to \pi$\, case \cite{Flynn:2015mha}:\\ [5pt]
$t_+ = (m_{B^+}+m_{\pi^+})^2$, \, $t_- = (m_{B^+}- m_{\pi^+})^2$, \\[2pt]  $t_0 = (m_{B^+}+m_{\pi^+})(\sqrt{m_{B^+}}-\sqrt{m_{\pi^+}})^2$, \\[3pt]
$b_+^{(0)} = 0.42$, \, $b_+^{(1)} = -1.46 \,b_+^{(0)}$, \, $b_+^{(2)} = -4.7\,b_+^{(0)}$,  \\ $b_0^{(0)} = 0.516$, \, $b_0^{(1)} = -3.94 \,b_0^{(0)}$, \, $b_0^{(2)} = 0.7\,b_0^{(0)}$, \\ [3pt] $P_+(q^2) = 1-q^2/m_{B^*}^2$, \, $m_{B^*} = 5.325 \ {\rm GeV}$,  \\

\noindent
whereas for \,$B \to K$\, \cite{Bouchard:2013pna}:\\[5pt]
$t_+ = (m_{B^+}+m_{K^+})^2$, \, $t_- = (m_{B^+}- m_{K^+})^2$, \\[2pt]  $t_0 = (m_{B^+}+m_{K^+})(\sqrt{m_{B^+}}-\sqrt{m_{K^+}})^2$, \\[3pt]
$b_+^{(0)} = 0.432$, \, $b_+^{(1)} = -0.65$, \, $b_+^{(2)} = -0.97$, \\ $b_0^{(0)} = 0.550$,  \, $b_0^{(1)} = -1.89$, \, $b_0^{(2)} = 1.98$, \, $b_0^{(3)} = -0.02$, \\ [3pt]
$P_+(q^2) = 1-q^2/(m_{B^+}+\Delta_+^*)^2$, \, $\Delta_+^* = 0.04578 \ {\rm GeV}$.  \\

The resulting constraints on $B^+$  decays are much weaker than the corresponding bounds presented in \cite{Kuznetsov:2012ai}. This is due to the fact that the calculation in \cite{Kuznetsov:2012ai} assumed  $f_{+}(q^2) = f_0(q^2) = 1$. This assumption for the $B \to \pi$ and $B\to K$ form factors is quite far from their actual shape.

\vspace{5mm}
\centerline
{{(a)} \bf \ {\emph{Charged kaon decays}}}\vspace{3mm}

\noindent
Experimental constraints from searches for the processes $K^+\rightarrow \pi^+ e^\pm \mu^\mp $ yield
 \bea\label{limits122}
  &&  \ \ \  {N_{12}^{+\,(K^+\!,\pi^+)}}+(0.54 \ {\rm GeV}^2) \ {P_{12}^{+\,(K^+\!,\pi^+)}}\hspace{18.8mm}\text{\cite{Appel:2000tc}}\nn \\
&&\hspace{3mm} + \  (0.83 \ {\rm GeV}) \, \big(R^{+\,(K^+\!,\pi^+)}_{12}-R^{-\,(K^+\!,\pi^+)}_{12}\big) \nn\\
&&\hspace{3.5mm}  \lesssim \ \left[\,\frac{{\rm Br}(K^+\rightarrow \pi^+ e^+ \mu^- )}{5.2 \times 10^{-10}}\right] (32.1 \ \,{\rm TeV})^{-4} \ , \\[7pt]
   &&  \ \ \  {N_{21}^{+\,(K^+\!,\pi^+)}}+(0.54 \ {\rm GeV}^2) \ {P_{21}^{+\,(K^+\!,\pi^+)}} \hspace{18.8mm} \text{\cite{Sher:2005sp}}\nn\\
&& \hspace{3mm}- \  (0.83 \ {\rm GeV}) \, \big(R^{+\,(K^+\!,\pi^+)}_{21}+R^{-\,(K^+\!,\pi^+)}_{21}\big)\nn\\
&&\hspace{3.5mm}  \lesssim \ \left[\,\frac{{\rm Br}(K^+\rightarrow \pi^+ e^- \mu^+ )}{1.3 \times 10^{-11} }\right] (80.6 \ \,{\rm TeV})^{-4} \ .
\eea

\vspace{3mm}
\centerline
{{(b)} \bf \ {\emph{Charged B meson decays}}}\vspace{3mm}

 \noindent
The experimental bounds on the decays $B^+\rightarrow \pi^+ e^\pm \mu^\mp$,  $B^+\rightarrow K^+ e^\pm \mu^\mp$ and  $B^+\rightarrow K^+ \mu^\pm \tau^\mp$ give

\ 
\vspace{-4mm}
\bea
  &&  \ \ \ {N_{12}^{+\,(B^+\!,\pi^+)}}+(138 \ {\rm GeV}^2) \ {P_{12}^{+\,(B^+\!,\pi^+)}}\hspace{20.5mm}\text{\cite{Aubert:2007mm}}  \nn\\
&& \hspace{2.5mm}+ \  (0.76\ {\rm GeV})  \ \big(R^{+\,(B^+\!,\pi^+)}_{12}-R^{-\,(B^+\!,\pi^+)}_{12}\big) \nn\\
&&\hspace{3mm}  \lesssim \ \left[\,\frac{{\rm Br}(B^+\rightarrow \pi^+ e^+ \mu^-)}{9.2 \times 10^{-8}}\right] (13.5 \ {\rm TeV})^{-4}\ ,\\[7pt]
     &&  \ \ \ {N_{21}^{+\,(B^+\!,\pi^+)}}+(138 \ {\rm GeV}^2) \ {P_{21}^{+\,(B^+\!,\pi^+)}}\hspace{20.5mm}\text{\cite{Aubert:2007mm}}\nn\\
&& \hspace{2.5mm}- \  (0.76\ {\rm GeV})  \ \big(R^{+\,(B^+\!,\pi^+)}_{21}+R^{-\,(B^+\!,\pi^+)}_{21}\big) \nn\\
&&\hspace{3mm}  \lesssim \ \left[\,\frac{{\rm Br}(B^+\rightarrow \pi^+ e^- \mu^+)}{9.2 \times 10^{-8}}\right] (13.5 \ {\rm TeV})^{-4}\ ,\\[7pt]
  && \ \ \ {N_{12}^{+\,(B^+\!,K^+)}}+(109 \ {\rm GeV}^2) \ {P_{12}^{+\,(B^+\!,K^+)}}\hspace{19mm}\text{\cite{Aubert:2006vb}}  \nn\\
&& \hspace{2.5mm}+ \  (1.0 \ {\rm GeV}) \,\big(R^{+\,(B^+\!,K^+)}_{12}-R^{-\,(B^+\!,K^+)}_{12}\big) \nn\\
&&\hspace{3mm}  \lesssim \ \left[\,\frac{{\rm Br}(B^+\rightarrow K^+ e^+ \mu^- )}{9.1 \times 10^{-8}}\right] (16.2  \ {\rm TeV})^{-4}\ ,\\[7pt]
     && \ \ \ {N_{21}^{+\,(B^+\!,K^+)}}+(109 \ {\rm GeV}^2) \ {P_{21}^{+\,(B^+\!,K^+)}}\hspace{19mm}\text{\cite{Aubert:2006vb}}\nn\\
&& \hspace{2.5mm}- \  (1.0 \ {\rm GeV}) \,\big(R^{+\,(B^+\!,K^+)}_{21}+R^{-\,(B^+\!,K^+)}_{21}\big) \nn\\
&&\hspace{3mm}  \lesssim \ \left[\,\frac{{\rm Br}(B^+\rightarrow K^+ e^- \mu^+)}{1.3 \times 10^{-7}}\right] (14.9 \ {\rm TeV})^{-4}\ ,\\[7pt]
  &&  \ \ \ {N_{23}^{+\,(B^+\!,K^+)}}+(96 \ {\rm GeV}^2) \ {P_{23}^{+\,(B^+\!,K^+)}}\hspace{20.8mm}\text{\cite{Aubert:2007rn}} \nn\\
&& \hspace{2.5mm}+ \  (10.3 \ {\rm GeV}) \,\big(R^{+\,(B^+\!,K^+)}_{23}-1.2\,R^{-\,(B^+\!,K^+)}_{23}\big) \nn\\
&&\hspace{3mm}  \lesssim \ \left[\,\frac{{\rm Br}(B^+\rightarrow K^+ \mu^+ \tau^-)}{7.7 \times 10^{-5} }\right] (2.7 \ {\rm TeV})^{-4}\ ,\\[7pt]
  && \ \ \ {N_{32}^{+\,(B^+\!,K^+)}}+(96 \ {\rm GeV}^2) \ {P_{32}^{+\,(B^+\!,K^+)}}\hspace{20.8mm}\text{\cite{Aubert:2007rn}} \nn\\
&& \hspace{2.5mm}- \  (10.3 \ {\rm GeV}) \,\big(R^{+\,(B^+\!,K^+)}_{23}+1.2\,R^{-\,(B^+\!,K^+)}_{23}\big) \nn\\
&&\hspace{3mm}  \lesssim \ \left[\,\frac{{\rm Br}(B^+\rightarrow K^+ \mu^- \tau^+)}{7.7 \times 10^{-5} }\right] (2.7 \ {\rm TeV})^{-4}\ .
\eea
\\

\vspace{0mm}
\centerline
{{(4)} \bf \ {\emph{Tau  decays}}}\vspace{3mm}

The leptoquark contribution to the rate of $\tau$ decays to a pseudoscalar meson and a lepton, neglecting the mass of the lepton in the final state,  is
\bea
\label{c48}
 &&\Gamma(\tau^-\rightarrow \mathcal{M}' \,l_i^-)_X \ = \ \frac{m_\tau^3 f^2_{{\mathcal{M}'}}}{128 \, \pi} \, T_i \left(1-\frac{m_{\mathcal{M}'}^2}{m_\tau^2}\right)^{2} , \ \ \ \ \ \ \ 
   \eea
   where
   \bea
  && T_i\  \equiv \ {\bigg|\sum_{\alpha}\frac{t_{i(\alpha)}^{L}}{M_\alpha^2}\bigg|^2}+ \, {\bigg|\sum_{\alpha}\frac{t_{i(\alpha)}^{R}}{M_\alpha^2}\bigg|^2}     \ , \nn\\
      &&t_{i(\alpha)}^{L\,(\pi^0)}\, \equiv \ f^{Rd}_{13(\alpha)} f^{Rd\,*}_{1i(\alpha)} + \frac{2m_{\pi^0}^2Q}{m_\tau (m_{d}+m_u)}f^{Ld}_{13(\alpha)} f^{Rd\,*}_{1i(\alpha)}  \,\ , \nn\\
 &&t_{i(\alpha)}^{L\,(K^0_S)} \equiv \ \frac1{\sqrt2}\bigg[f^{Rd}_{13(\alpha)} f^{Rd\,*}_{2i(\alpha)} + \frac{2m_{{K^0_S}}^2Q}{m_\tau(m_{s}+m_{d})}f^{Ld}_{13(\alpha)} f^{Rd\,*}_{2i(\alpha)}\bigg]\nn\\
 &&\hspace{16mm}  - \  \left(1\leftrightarrow 2\right)\ , \nn\\[4pt]
  &&t_{i(\alpha)}^{R} \equiv \ t_{i(\alpha)}^{L}\left(L\leftrightarrow R\right)  
  \eea
  and $f_{\pi^0} = f_{\pi^+}/\sqrt2$. 
  The bounds are
\bea
 && \ \ \  T_1^{(\pi^0)} \ \lesssim \ \left[\,\frac{{\rm Br}(\tau^- \rightarrow  \pi^0\,e^-)}{8.0 \times 10^{-8}}\right] \,(5.0 \ \,{\rm TeV})^{-4} \ ,\hspace{9.0mm}\text{\cite{Miyazaki:2007jp}}\nn\\
 &&\\[4pt]
  &&  \ \ \  T_2^{(\pi^0)} \ \lesssim \ \left[\,\frac{{\rm Br}(\tau^- \rightarrow \pi^0\,\mu^- )}{1.1 \times 10^{-7} }\right] \,(4.7 \ \,{\rm TeV})^{-4}\ ,\hspace{8.5mm}\text{\cite{Aubert:2006cz}}\nn\\
 &&\\[4pt]
 &&  \ \ \  T_1^{(K^0_S)} \, \lesssim \ \left[\,\frac{{\rm Br}(\tau^- \rightarrow K^0_S\, e^-)}{2.6 \times 10^{-8} }\right]\, (8.4 \ \,{\rm TeV})^{-4}\ ,\hspace{7mm}\text{\cite{Miyazaki:2010qb}} \nn\\
 &&\\[4pt]
  && \ \ \  T_2^{(K^0_S)} \, \lesssim \ \left[\,\frac{{\rm Br}(\tau^- \rightarrow K^0_S \,\mu^- )}{2.3 \times 10^{-8}}\right] \,(8.6 \ \,{\rm TeV})^{-4} \ .\hspace{7mm}\text{\cite{Miyazaki:2010qb}} \nn\\
   &&
   \eea
   
There are also constraints from $\tau$ decays to a vector meson and a lepton, e.g. $\tau^- \to \phi\ l_i^-$. The resulting leptoquark contribution to the rate is given by Eq.\,(\ref{c48}) with
   \bea
    &&t_{i(\alpha)}^{L\,(\phi)}\equiv  \sqrt{1+\frac{2m_{\phi}^2}{m_\tau^2}}\Big(f^{Rd}_{23(\alpha)} f^{Rd\,*}_{2i(\alpha)} + \frac{m_{\phi}^2\,Q}{m_\tau m_s}f^{Ld}_{23(\alpha)} f^{Rd\,*}_{2i(\alpha)} \Big)   ,\nn\\[4pt]
     &&t_{i(\alpha)}^{R\,(\phi)} \equiv \ t_{i(\alpha)}^{L\,(\phi)}\left(L\leftrightarrow R\right)  .
    \eea
    \vspace{0mm}
    
   \noindent 
The experimental bounds yield
   \bea
  &&  \ \ \  T_1^{(\phi)} \, \lesssim \ \left[\,\frac{{\rm Br}(\tau^- \rightarrow \phi \,e^- )}{3.1 \times 10^{-8}}\right] \,(9.6 \ \,{\rm TeV})^{-4} \ ,\hspace{12.5mm}\text{\cite{Miyazaki:2011xe}}\nn\\
     &&\\[4pt]
  &&  \ \ \  T_2^{(\phi)} \, \lesssim \ \left[\,\frac{{\rm Br}(\tau^- \rightarrow \phi \,\mu^- )}{8.4 \times 10^{-8}}\right] \,(7.5 \ \,{\rm TeV})^{-4} \ .\hspace{12.5mm}\text{\cite{Miyazaki:2011xe}}\nn\\
\eea

\vspace{3mm}
\centerline
{{(5)} \bf \ {\emph{Radiative charged lepton decay}}}\vspace{3mm}

The vector leptoquark contribution to the process ${\l_i \to l_j \gamma}$ is induced at the loop level. Unlike for scalar leptoquarks, in the case of vector leptoquarks this effect cannot be computed in the general case, since the result is infinite and requires arbitrary subtractions that are well defined only in a UV complete model. We parameterize our ignorance of this UV completion with the coefficients $c_{LR}$ and $c_{RL}$,
\bea\label{mueg}
\Gamma(l_i^+ \!\to l_j^+ \gamma)_X\,  &&=  \frac{e^2m_i^5}{4096\,\pi^5}\Bigg[ \bigg|\sum_{\alpha; \, k}\frac{f^{Ld}_{kj(\alpha)}f^{Ld*}_{ki(\alpha)}}{M_{\alpha}^2}\bigg|^2 \!\!+ (L \leftrightarrow R)\Bigg] \nn\\[2pt]
&&\hspace{-23mm} +\ \,  \frac{e^2m_i^3m_b^2}{4096\,\pi^5}  \Bigg[ \, c^2_{LR}\ \bigg|\sum_{\alpha}\frac{f^{Ld}_{3j(\alpha)}f^{Rd*}_{3i(\alpha)}}{M_{\alpha}^2}\bigg|^2 \!\!+ (L \leftrightarrow R)\Bigg]+ \ .\,.\,. \ \,,\nn\\
 \eea
 where $k=1,2,3$ and we expect $c_{LR}$ and $c_{RL}$ to be $\mathcal{O}(1)$, with their  values  dependent on the UV details of the model. The ellipsis denotes interference and mass-suppressed terms.
 
 If the matrices $f_{ij}$ are proportional to unitary matrices, the terms in the first line of Eq.~(\ref{mueg}) vanish. The experimental bounds, neglecting higher order terms, become
\bea\label{300}
&& \ \ \ c^2_{LR}\ \bigg|\sum_{\alpha}\frac{f^{Ld}_{31(\alpha)}f^{Rd*}_{32(\alpha)}}{M_{\alpha}^2}\bigg|^2 +  \,c^2_{RL}\ \bigg|\sum_{\alpha}\frac{f^{Rd}_{31(\alpha)}f^{Ld*}_{32(\alpha)}}{M_{\alpha}^2}\bigg|^2 \nn\\
&& \hspace{3mm}\lesssim \ \left[\,\frac{{\rm Br}(\mu \to e \,\gamma)}{4.2 \times 10^{-13}}\right] \,(332 \ {\rm TeV} )^{-4} \ , \ \ \ \ \ \text{\cite{TheMEG:2016wtm}}\\[10pt]
&& \ \ \ c^2_{LR}\ \bigg|\sum_{\alpha}\frac{f^{Ld}_{31(\alpha)}f^{Rd*}_{33(\alpha)}}{M_{\alpha}^2}\bigg|^2 +  \,c^2_{RL}\ \bigg|\sum_{\alpha}\frac{f^{Rd}_{31(\alpha)}f^{Ld*}_{33(\alpha)}}{M_{\alpha}^2}\bigg|^2 \nn\\
&& \hspace{3mm}\lesssim \ \left[\,\frac{{\rm Br}(\tau \to e \,\gamma)}{3.3 \times 10^{-8}}\right] \,(3.1 \ {\rm TeV} )^{-4} \ , \ \ \ \ \, \ \text{\cite{Aubert:2009ag}}\\[10pt]
&& \ \ \ c^2_{LR}\ \bigg|\sum_{\alpha}\frac{f^{Ld}_{32(\alpha)}f^{Rd*}_{33(\alpha)}}{M_{\alpha}^2}\bigg|^2 +  \,c^2_{RL}\ \bigg|\sum_{\alpha}\frac{f^{Rd}_{32(\alpha)}f^{Ld*}_{33(\alpha)}}{M_{\alpha}^2}\bigg|^2 \nn\\
&& \hspace{3mm}\lesssim \ \left[\,\frac{{\rm Br}(\tau \to \mu \,\gamma)}{4.4 \times 10^{-8}}\right] \,(2.9 \ {\rm TeV} )^{-4} \ . \ \ \ \ \ \, \text{\cite{Aubert:2009ag}}
\eea
In our model the leading order terms contributing to  $l_i^+ \!\to l_j^+ \gamma$ are $\mathcal{O}(m_b^2/M_{X_L}^2)$ and the resulting constraints  are negligible compared to tree-level bounds.\\

\vspace{3mm}
\centerline
{{(6)} \bf {\emph{{\boldsymbol{\ $l_i^+ \!\to l_j^+$}} \,{\bf\emph{conversion}}}}}\vspace{3mm}

The effective Hamiltonian for the ${l_i^+\! \to l_j^+}$ conversion consists of the dipole transition part corresponding to  ${\l_i^+ \to l_j^+ \gamma}$ and terms arising from integrating out the heavy vector leptoquarks, i.e.
\bea\label{muecon}
\mathcal{H}^{\rm eff}_{l_i^+\to\, l_j^+}\,&&= \frac{e\,m_b}{16\,\pi^2}\sum_\alpha \frac{c_{LR}}{M_{\alpha}^2} \ {f^{Ld}_{3j(\alpha)}f^{Rd*}_{3i(\alpha)}} \ \overline{l}_{jR}\, \sigma_{\mu\nu}\, l_{iL} F^{\mu\nu} \nn\\
&& \hspace{0mm}+ \ \frac{e\,m_i}{16\,\pi^2}\sum_{\alpha, \,k} \frac{1}{M_{\alpha}^2} \ {f^{Ld}_{kj(\alpha)}f^{Ld*}_{ki(\alpha)}} \ \overline{l}_{jR}\, \sigma_{\mu\nu}\, l_{iL} F^{\mu\nu} \nn\\
&& \hspace{0mm}+ \ \sum_{\alpha, \,m} \frac{1}{M_{\alpha}^2} \Big[ {f^{Ld}_{mj(\alpha)}f^{Ld*}_{mi(\alpha)}}\big(\overline{l}_{jL}\gamma^\mu l_{iL}\big)\big(\overline{d}^{\,k}_{L} \gamma_\mu d^{\,k}_L\big)\nn\\
&&\hspace{14mm}-  \ 2\,Q \,{f^{Ld}_{mj(\alpha)}f^{Rd*}_{mi(\alpha)}}\big(\overline{l}_{jL} l_{iR}\big)\big(\overline{d}^{\,k}_{R}\,  d^{\,k}_L\big) \Big]\nn\\
&& +\  (L \leftrightarrow R)\, + \, \ \dots \ ,
\eea
where $m=1,2$. The steps required to match the effective Hamiltonian (\ref{muecon}) to the Hamiltonian at the nucleon level and compute the conversion rate are provided in \cite{Kitano:2002mt,Cirigliano:2009bz}. 

The tightest experimental constraint from $l_i^+ \!- l_j^+$ conversion arises from $\mu - e$ conversion on gold \cite{Bertl:2006up}. Since the resulting bound on the dipole transition contribution is  less restrictive than the constraint from $\mu \to e \,\gamma$  in Eq.~(\ref{300}), we concentrate only on the second part of the Hamiltonian (\ref{muecon}).
Following \cite{Kitano:2002mt},  the $\mu - e$ conversion rate is then given by
\bea\label{capG}
\Gamma(\mu \to e) &= &  m_\mu^5\,\big| \tilde{g}_{LV}^{(p)} V_{p}+\tilde{g}_{LV}^{(n)} V_{n} + \tilde{g}_{LS}^{(p)} S_{p}+\tilde{g}_{LS}^{(n)} S_{n} \big|^2 \nn\\[2pt]
&&+ \ (L \leftrightarrow R) \ , \ \ \ \ \ \ \ 
\eea
where 
\bea
&&\tilde{g}_{LV}^{(p)} = \tfrac12\,\tilde{g}_{LV}^{(n)} =  \sum_{\alpha} \frac{1}{M_{\alpha}^2}  {f^{Ld}_{12(\alpha)}f^{Ld*}_{11(\alpha)}} \ ,\\
&&\tilde{g}_{LS}^{(p)}= -2\,Q  \sum_{\alpha} \frac{1}{M_{\alpha}^2} \left[4.3\,{f^{Ld}_{12(\alpha)}f^{Rd*}_{11(\alpha)}}+ 2.5\,{f^{Ld}_{22(\alpha)}f^{Rd*}_{21(\alpha)}}\right], \nn\\
&&\tilde{g}_{LS}^{(n)}= -2\,Q  \sum_{\alpha} \frac{1}{M_{\alpha}^2} \left[5.1\,{f^{Ld}_{12(\alpha)}f^{Rd*}_{11(\alpha)}}+ 2.5\,{f^{Ld}_{22(\alpha)}f^{Rd*}_{21(\alpha)}}\right] , \nn
\eea
with  similar relations obtained upon switching $(L\leftrightarrow R)$. The numerical coefficients were adopted from \cite{Kosmas:2001mv}. For the $^{197}_{\ \, 79}{\rm Au}$ nucleus, which provides the most stringent   bound, the parameters in Eq.\,(\ref{capG}) are,
\bea
V_p = 0.0974 \ , \ \ V_n=0.146 \ ,  \ \ S_p = 0.0614  \ , \ \ S_n = 0.0918 \ ,\nn
\eea
and they are the result of the calculation using ``method 1'' in Sec. III A of \cite{Kitano:2002mt}. The best bound on $\mu-e$ conversion is \cite{Bertl:2006up}
\bea\label{BAU}
 \frac{\Gamma(\mu \to e \ {\rm in \ Au})}{\Gamma(\mu \ {\rm capture \ in \  Au})}  <  7 \times 10^{-13} \ . \ \ \
\eea

The constraints on general $(3,1)_{2/3}$ leptoquark models are derived by inserting Eq.\,(\ref{capG}) into (\ref{BAU}) and using the total $\mu^-$ capture rate in $^{197}_{\ \, 79}{\rm Au}$,  $\Gamma(\mu \ {\rm capture \ in \  Au}) = 8.6 \times 10^{-18}  \ {\rm GeV}$  \cite{Suzuki:1987jf}. In the case of our model, with just LH leptoquark couplings, the constraint simplifies to
\bea\label{400}
&&  \bigg|\sum_{\alpha}\frac{f^{Ld}_{12(\alpha)}f^{Ld*}_{11(\alpha)}}{M_{\alpha}^2}\bigg|^{-1/2} \!\!\gtrsim  \ 762 \ {\rm TeV} \ . \ \ \ \ \ \ \ \ 
\eea

Finally, let us note that  the 
bounds on generic leptoquark models were considered in \cite{Valencia:1994cj,Smirnov:2007hv,Smirnov:2008zzb,Carpentier:2010ue,Kuznetsov:2012ai,Smirnov:2018ske}. Our formulae reproduce those results up to the difference in the adopted values of quark masses, meson decay constants and form factors used.\\

\section{\ Flavor constraints:\,   ${\rm \bf SU}{\bf (4)}_{{\textbf{\textit{{L}}}}} \times {\rm \bf SU}{\bf (4)}_{{\textbf{\textit{{R}}}}}$\,  model }\label{aapp2}

\noindent
In our model $X^{(1)} \equiv X_{1}$ and  $X^{(2)} \equiv X_{2}$\, given by Eq.~(\ref{xmatr}); therefore the coefficients in Eq.~(\ref{glag}) are
\bea\label{fijf}
\begin{aligned}
&f^{Lu}_{ij(1)} \equiv \frac{g_L\cos\theta_4}{\sqrt2}L^u_{ij} \ , \ \ \ \ \ \ f^{Ld}_{ij(1)} \equiv \frac{g_L\cos\theta_4}{\sqrt2}L^d_{ij} \ , \ \ \ \ \ \ \ \ \ \ \\
& f^{Ru}_{ij(1)} \equiv \frac{g_R\sin\theta_4}{\sqrt2}R^u_{ij} \ , \ \ \ \ \ \ f^{Rd}_{ij(1)} \equiv \frac{g_R\sin\theta_4}{\sqrt2}R^d_{ij} \ , \\
&f^{Ru}_{ij(2)} \equiv \frac{g_R\cos\theta_4}{\sqrt2}R^u_{ij} \ , \ \ \ \ \ \ f^{Rd}_{ij(2)} \equiv \frac{g_R\cos\theta_4}{\sqrt2}R^d_{ij} \ , \\
&f^{Lu}_{ij(2)} \equiv -\frac{g_L\sin\theta_4}{\sqrt2}L^u_{ij} \ , \ \ \ \ f^{Ld}_{ij(2)} \equiv - \frac{g_L\sin\theta_4}{\sqrt2}L^d_{ij} \ .\ \ 
\end{aligned}
\eea

Constraints  on the model parameters are obtained by substituting the expressions in Eq.\,(\ref{fijf}) into the bounds derived in App.~\ref{aapp1}. 
In the limit 
$
v_R \gg v_L$ and $v_R \gg v_\Sigma
$,
for which \,$\sin\theta_4 \simeq 0$, $X_1 = X_L$ and $X_2 = X_R$, one arrives at the constraints listed below. The numbering scheme indicates which equation in App.~\ref{aapp1} a given constraint originated from.

{\bf {\emph{$\boldsymbol{K_L^0}$ decays}}}\vspace{1mm}
 \begin{align}
&\hspace{0mm} \frac{M_{X_L}}{g_L\sqrt{|{\rm Re}(L^d_{11}L^{d*}_{21})|}} \gtrsim \ 21.2 \ {\rm TeV} \ , \tag{D13}\\[3pt]
 & \frac{M_{X_L}}{g_L\sqrt{|L^d_{11}L^{d*}_{22}+L^d_{21}L^{d*}_{12}|}} \gtrsim \ 225\ {\rm TeV} \ , \ \ \ \ \ \ \ \ \  \tag{D14}\\[3pt]
  &\hspace{0mm}\frac{M_{X_L}}{g_L\sqrt{|{\rm Re}(L^d_{12}L^{d*}_{22})|}} \gtrsim \ 51.0 \ {\rm TeV} \ . \ \ \ \ \ \ \ \tag{D15}
  \end{align}
\vspace{2mm}

  {\bf {\emph{$\boldsymbol{B^0}$ decays}}}\vspace{1mm}
\begin{align}
&\hspace{0mm}  \frac{M_{X_L}}{g_L\sqrt{|L^d_{11}L^{d*}_{31}|}} \gtrsim \ 0.24 \ {\rm TeV} \ , \tag{D16}\\[3pt]
&  \frac{M_{X_L}}{g_L\sqrt[4]{|L^d_{11}L^{d*}_{32}|^2 + |L^d_{12}L^{d*}_{31}|^2}} \gtrsim \ 8.9 \ {\rm TeV} \ , \ \ \ \ \tag{D17}\\[5pt]
  &\hspace{0mm}   \frac{M_{X_L}}{g_L\sqrt{|L^d_{12}L^{d*}_{32}|}} \gtrsim \ 10.7 \ {\rm TeV}\tag{D18} \ ,\\[3pt]
  &  \frac{M_{X_L}}{g_L\sqrt[4]{|L^d_{11}L^{d*}_{33}|^2+|L^d_{13}L^{d*}_{31}|^2}}\gtrsim \ 2.6 \ {\rm TeV} \ ,\tag{D19} \\[3pt]
  &  \frac{M_{X_L}}{g_L\sqrt[4]{|L^d_{12}L^{d*}_{33}|^2+|L^d_{13}L^{d*}_{32}|^2}}\gtrsim \ 2.8 \ {\rm TeV} \ ,\tag{D20} \\[3pt]
&\hspace{0mm}  \frac{M_{X_L}}{g_L\sqrt{|L^d_{13}L^{d*}_{33}|}} \gtrsim \ 1.0 \ {\rm TeV} \ .\tag{D21}
\end{align}
\vspace{2mm}

{\bf {\emph{$\boldsymbol{B_s^0}$ decays}}}\vspace{1mm}
\begin{align}
&\hspace{0mm} \frac{M_{X_L}}{g_L\sqrt{|L^d_{21}L^{d*}_{31}|}} \gtrsim \ 0.2 \ {\rm TeV} \ ,\tag{D22} \\[3pt]
&  \frac{M_{X_L}}{g_L\sqrt[4]{|L^d_{21}L^{d*}_{32}|^2+|L^d_{22}L^{d*}_{31}|^2}} \gtrsim \ 6.4 \ {\rm TeV} \ , \ \ \ \ \tag{D23}\\[3pt]
& \hspace{0mm} \frac{M_{X_L}}{g_L\sqrt{|L^d_{22}L^{d*}_{32}|}} \gtrsim \ 6.0 \ {\rm TeV} \ , \tag{D24}\\[3pt]
&\hspace{0mm}  \frac{M_{X_L}}{g_L\sqrt{|L^d_{23}L^{d*}_{33}|}} \gtrsim \ 0.7 \ {\rm TeV} \ .\tag{D25}
  \end{align}
\vspace{2mm}

{\bf {\emph{$\boldsymbol{\pi^+}$ decays}}}\vspace{1mm}
 \begin{align}
 &&\frac{M_{X_L}}{g_L\sqrt{\,\big|{\rm Re}\!\left[{L^d_{11}(V L^{d})^*_{11}}-4.3\,{L^d_{12}(V L^{d})^*_{12}}\right]\!\big|}} \gtrsim \ 2.8 \ {\rm TeV} \ .  \tag{D32}
  \end{align}\vspace{2mm}

{\bf {\emph{$\boldsymbol{K^+}$ decays}}}\vspace{1mm}
 \begin{align}
 &\frac{M_{X_L}}{g_L\sqrt{\,\big|{\rm Re}\!\left[{L^d_{11}(V L^{d})^*_{21}}-4.3\,{L^d_{12}(V L^{d})^*_{22}}\right]\!\big|}} \gtrsim \ 2.2 \ {\rm TeV} \, ,  \tag{D33}
 \end{align}
 \begin{align}
  &\hspace{-20mm}\frac{M_{X_L}}{g_L\sqrt{|L^d_{11}L^{d*}_{22}|}} \gtrsim \ 27.0 \ {\rm TeV} \ ,\tag{D40}\\[3pt]
 & \hspace{-20mm}\frac{M_{X_L}}{g_L\sqrt{|L^d_{12}L^{d*}_{21}|}} \gtrsim \ 67.8 \ {\rm TeV} \ .\tag{D41}
 \end{align}
\vspace{1mm}

{\bf {\emph{$\boldsymbol{B^+}$ decays}}}\vspace{0mm}
 \begin{align}
        &   \hspace{-10mm} \frac{M_{X_L}}{g_L\sqrt{|L^d_{11}L^{d*}_{32}|}}  \gtrsim \ 11.4 \ {\rm TeV} \ ,\tag{D42} \\[9pt]
                &   \hspace{-10mm} \frac{M_{X_L}}{g_L\sqrt{|L^d_{12}L^{d*}_{31}|}} \gtrsim \ 11.4 \ {\rm TeV} \ ,\tag{D43} \\[9pt]
  & \hspace{-10mm}\frac{M_{X_L}}{g_L\sqrt{|L^d_{21}L^{d*}_{32}|}} \gtrsim \ 13.6 \ {\rm TeV} \ ,\tag{D44} \ \ \ \ \ \ \ \ \ \ \ \\[9pt]
  & \hspace{-10mm}\frac{M_{X_L}}{g_L\sqrt{|L^d_{22}L^{d*}_{31}|}} \gtrsim \ 12.5 \ {\rm TeV} \  , \tag{D45}\\[9pt]
  &   \hspace{-10mm}    \frac{M_{X_L}}{g_L\sqrt{|L^d_{22}L^{d*}_{33}|}} \gtrsim \ 2.3 \ {\rm TeV}\ , \ \ \ \ \tag{D46}\\[9pt]
    &   \hspace{-10mm}   \frac{M_{X_L}}{g_L\sqrt{|L^d_{23}L^{d*}_{32}|}}\gtrsim \ 2.3 \ {\rm TeV} \ . \ \ \ \ \tag{D47}
  \end{align}

{\bf {\emph{$\boldsymbol{\tau}$ decays}}}\vspace{1mm}
\begin{align}
& \hspace{0mm}  \frac{M_{X_L}}{g_L\sqrt{|L^d_{11}L^{d*}_{13}|}} \gtrsim \ 3.6 \ {\rm TeV} \ , \tag{D50}\\[3pt]
& \hspace{0mm}\frac{M_{X_L}}{g_L\sqrt{|L^d_{12}L^{d*}_{13}|}} \gtrsim \ 3.3 \ {\rm TeV} \ , \tag{D51} \\[3pt]
&\frac{M_{X_L}}{g_L\sqrt{\big|L^d_{21}L^{d*}_{13}-L^d_{11}L^{d*}_{23}\big|}} \gtrsim \ 5.0 \ {\rm TeV}  \ ,  \tag{D52}\\[3pt]
&  \frac{M_{X_L}}{g_L\sqrt{\big|L^d_{22}L^{d*}_{13}-L^d_{12}L^{d*}_{23}\big|}} \gtrsim  \ 5.1 \ {\rm TeV} \ , \tag{D53}\\[3pt]
& \hspace{0mm}  \frac{M_{X_L}}{g_L\sqrt{|L^d_{21}L^{d*}_{23}|}} \gtrsim \ 6.8 \ {\rm TeV} \ , \tag{D55}\\[3pt]
& \hspace{0mm}\frac{M_{X_L}}{g_L\sqrt{|L^d_{22}L^{d*}_{23}|}} \gtrsim \ 5.3 \ {\rm TeV} \ . \tag{D56} 
  \end{align}
\vspace{2mm}

{\bf {\emph{$\boldsymbol{\mu - e}$ \ conversion}}}
\begin{align}
 \hspace{-23mm}  \frac{M_{X_L}}{g_L\sqrt{|L^d_{12}L^{d*}_{11}|}} \gtrsim \ 539 \ {\rm TeV} \ .\tag{D65}
  \end{align}

\bibliography{SU4}

\end{document}